%
%
%
\input epsf
\def\unredoffs{} \def\redoffs{\voffset=-.31truein\hoffset=-.59truein}
\def\speclscape{\special{ps: landscape}}
%
%
%
%
\newbox\leftpage \newdimen\fullhsize \newdimen\hstitle \newdimen\hsbody
\tolerance=1000\hfuzz=2pt
\catcode`\@=11 
\def\bigans{b }
\def\answ{b }
\ifx\answ\bigans\message{(This will come out unreduced.}
\magnification=1200\unredoffs\baselineskip=16pt plus 2pt minus 1pt
\hsbody=\hsize \hstitle=\hsize 
\else\message{(This will be reduced.} \let\l@r=L
\magnification=1000\baselineskip=16pt plus 2pt minus 1pt \vsize=7truein
\redoffs \hstitle=8truein\hsbody=4.75truein\fullhsize=10truein\hsize=\hsbody
\output={\ifnum\pageno=0 
  \shipout\vbox{\speclscape{\hsize\fullhsize\makeheadline}
    \hbox to \fullhsize{\hfill\pagebody\hfill}}\advancepageno
  \else
  \almostshipout{\leftline{\vbox{\pagebody\makefootline}}}\advancepageno
  \fi}
\def\almostshipout#1{\if L\l@r \count1=1 \message{[\the\count0.\the\count1]}
      \global\setbox\leftpage=#1 \global\let\l@r=R
 \else \count1=2
  \shipout\vbox{\speclscape{\hsize\fullhsize\makeheadline}
      \hbox to\fullhsize{\box\leftpage\hfil#1}}  \global\let\l@r=L\fi}
\fi
%
\newcount\yearltd\yearltd=\year\advance\yearltd by -1900

\def\Title#1#2{\nopagenumbers\abstractfont\hsize=\hstitle\rightline{#1}%
\vskip 1in\centerline{\titlefont #2}\abstractfont\vskip .5in\pageno=0}
\def\Date#1{\vfill\leftline{#1}\tenpoint\supereject\global\hsize=\hsbody%
\footline={\hss\tenrm\folio\hss}}
%

\def\draftmode{\message{ DRAFTMODE }\def\draftdate{{\rm preliminary draft:
\number\month/\number\day/\number\yearltd\ \ \hourmin}}%
\headline={\hfil\draftdate}\writelabels\baselineskip=20pt plus 2pt minus 2pt
 {\count255=\time\divide\count255 by 60 \xdef\hourmin{\number\count255}
  \multiply\count255 by-60\advance\count255 by\time
  \xdef\hourmin{\hourmin:\ifnum\count255<10 0\fi\the\count255}}}
\def\nolabels{\def\wrlabeL##1{}\def\eqlabeL##1{}\def\reflabeL##1{}}
\def\writelabels{\def\wrlabeL##1{\leavevmode\vadjust{\rlap{\smash%
{\line{{\escapechar=` \hfill\rlap{\sevenrm\hskip.03in\string##1}}}}}}}%
\def\eqlabeL##1{{\escapechar-1\rlap{\sevenrm\hskip.05in\string##1}}}%
\def\reflabeL##1{\noexpand\llap{\noexpand\sevenrm\string\string\string##1}}}
\nolabels
%
\global\newcount\secno \global\secno=0
\global\newcount\meqno \global\meqno=1
\def\newsec#1{\global\advance\secno by1\message{(\the\secno. #1)}
\global\subsecno=0\eqnres@t\noindent{\bf\the\secno. #1}
\writetoca{{\secsym} {#1}}\par\nobreak\medskip\nobreak}
\def\eqnres@t{\xdef\secsym{\the\secno.}\global\meqno=1\bigbreak\bigskip}
\def\sequentialequations{\def\eqnres@t{\bigbreak}}\xdef\secsym{}
\global\newcount\subsecno \global\subsecno=0
\def\subsec#1{\global\advance\subsecno by1\message{(\secsym\the\subsecno.
#1)}
\ifnum\lastpenalty>9000\else\bigbreak\fi
\noindent{\it\secsym\the\subsecno. #1}\writetoca{\string\quad
{\secsym\the\subsecno.} {#1}}\par\nobreak\medskip\nobreak}
\def\appendix#1#2{\global\meqno=1\global\subsecno=0\xdef\secsym{\hbox{#1.}}
\bigbreak\bigskip\noindent{\bf Appendix #1. #2}\message{(#1. #2)}
\writetoca{Appendix {#1.} {#2}}\par\nobreak\medskip\nobreak}
%
%
\def\eqnn#1{\xdef #1{(\secsym\the\meqno)}\writedef{#1\leftbracket#1}%
\global\advance\meqno by1\wrlabeL#1}
\def\eqna#1{\xdef #1##1{\hbox{$(\secsym\the\meqno##1)$}}
\writedef{#1\numbersign1\leftbracket#1{\numbersign1}}%
\global\advance\meqno by1\wrlabeL{#1$\{\}$}}
\def\eqn#1#2{\xdef #1{(\secsym\the\meqno)}\writedef{#1\leftbracket#1}%
\global\advance\meqno by1$$#2\eqno#1\eqlabeL#1$$}
%
\newskip\footskip\footskip14pt plus 1pt minus 1pt 
\def\footnotefont{\ninepoint}\def\f@t#1{\footnotefont #1\@foot}
\def\f@@t{\baselineskip\footskip\bgroup\footnotefont\aftergroup\@foot\let\next}
\setbox\strutbox=\hbox{\vrule height9.5pt depth4.5pt width0pt}
\global\newcount\ftno \global\ftno=0
\def\foot{\global\advance\ftno by1\footnote{$^{\the\ftno}$}}
%
\newwrite\ftfile
\def\footend{\def\foot{\global\advance\ftno by1\chardef\wfile=\ftfile
$^{\the\ftno}$\ifnum\ftno=1\immediate\openout\ftfile=foots.tmp\fi%
\immediate\write\ftfile{\noexpand\smallskip%
\noexpand\item{f\the\ftno:\ }\pctsign}\findarg}%
\def\footatend{\vfill\eject\immediate\closeout\ftfile{\parindent=20pt
\centerline{\bf Footnotes}\nobreak\bigskip\input foots.tmp }}}
\def\footatend{}
%
%
\global\newcount\refno \global\refno=1
\newwrite\rfile
\def\ref{[\the\refno]\nref}
\def\nref#1{\xdef#1{[\the\refno]}\writedef{#1\leftbracket#1}%
\ifnum\refno=1\immediate\openout\rfile=refs.tmp\fi
\global\advance\refno by1\chardef\wfile=\rfile\immediate
\write\rfile{\noexpand\item{#1\ }\reflabeL{#1\hskip.31in}\pctsign}\findarg}
\def\findarg#1#{\begingroup\obeylines\newlinechar=`\^^M\pass@rg}
{\obeylines\gdef\pass@rg#1{\writ@line\relax #1^^M\hbox{}^^M}%
\gdef\writ@line#1^^M{\expandafter\toks0\expandafter{\striprel@x #1}%
\edef\next{\the\toks0}\ifx\next\em@rk\let\next=\endgroup\else\ifx\next\empty%
\else\immediate\write\wfile{\the\toks0}\fi\let\next=\writ@line\fi\next\relax}}
\def\striprel@x#1{} \def\em@rk{\hbox{}}
\def\lref{\begingroup\obeylines\lr@f}
\def\lr@f#1#2{\gdef#1{\ref#1{#2}}\endgroup\unskip}
\def\semi{;\hfil\break}
\def\addref#1{\immediate\write\rfile{\noexpand\item{}#1}} 
\def\startrefs#1{\immediate\openout\rfile=refs.tmp\refno=#1}
\def\xref{\expandafter\xr@f}\def\xr@f[#1]{#1}
\def\refs#1{\count255=1[\r@fs #1{\hbox{}}]}
\def\r@fs#1{\ifx\und@fined#1\message{reflabel \string#1 is undefined.}%
\nref#1{need to supply reference \string#1.}\fi%
\vphantom{\hphantom{#1}}\edef\next{#1}\ifx\next\em@rk\def\next{}%
\else\ifx\next#1\ifodd\count255\relax\xref#1\count255=0\fi%
\else#1\count255=1\fi\let\next=\r@fs\fi\next}
%

%
\newwrite\ffile\global\newcount\figno \global\figno=1
\def\fig{fig.~\the\figno\nfig}
\def\nfig#1{\xdef#1{fig.~\the\figno}%
\writedef{#1\leftbracket fig.\noexpand~\the\figno}%
\ifnum\figno=1\immediate\openout\ffile=figs.tmp\fi\chardef\wfile=\ffile%
\immediate\write\ffile{\noexpand\medskip\noexpand\item{Fig.\ \the\figno. }
\reflabeL{#1\hskip.55in}\pctsign}\global\advance\figno by1\findarg}
\def\xfig{\expandafter\xf@g}\def\xf@g fig.\penalty\@M\ {}
\def\figs#1{figs.~\f@gs #1{\hbox{}}}
\def\f@gs#1{\edef\next{#1}\ifx\next\em@rk\def\next{}\else
\ifx\next#1\xfig #1\else#1\fi\let\next=\f@gs\fi\next}
\newwrite\lfile
{\escapechar-1\xdef\pctsign{\string\%}\xdef\leftbracket{\string\{}
\xdef\rightbracket{\string\}}\xdef\numbersign{\string\#}}

\def\writestop{\def\writestoppt{\immediate\write\lfile{\string\pageno%
\the\pageno\string\startrefs\leftbracket\the\refno\rightbracket%
\string\def\string\secsym\leftbracket\secsym\rightbracket%
\string\secno\the\secno\string\meqno\the\meqno}\immediate\closeout\lfile}}
\def\writestoppt{}\def\writedef#1{}
\def\seclab#1{\xdef #1{\the\secno}\writedef{#1\leftbracket#1}\wrlabeL{#1=#1}}
\def\subseclab#1{\xdef #1{\secsym\the\subsecno}%
\writedef{#1\leftbracket#1}\wrlabeL{#1=#1}}
\newwrite\tfile \def\writetoca#1{}
\def\leaderfill{\leaders\hbox to 1em{\hss.\hss}\hfill}
\def\writetoc{\immediate\openout\tfile=toc.tmp
   \def\writetoca##1{{\edef\next{\write\tfile{\noindent ##1
   \string\leaderfill {\noexpand\number\pageno} \par}}\next}}}

%
\catcode`\@=12 
%
\edef\tfontsize{\ifx\answ\bigans scaled\magstep3\else scaled\magstep4\fi}
\font\titlerm=cmr10 \tfontsize \font\titlerms=cmr7 \tfontsize
\font\titlermss=cmr5 \tfontsize \font\titlei=cmmi10 \tfontsize
\font\titleis=cmmi7 \tfontsize \font\titleiss=cmmi5 \tfontsize
\font\titlesy=cmsy10 \tfontsize \font\titlesys=cmsy7 \tfontsize
\font\titlesyss=cmsy5 \tfontsize \font\titleit=cmti10 \tfontsize
\skewchar\titlei='177 \skewchar\titleis='177 \skewchar\titleiss='177
\skewchar\titlesy='60 \skewchar\titlesys='60 \skewchar\titlesyss='60
\def\titlefont{\def\rm{\fam0\titlerm}
\textfont0=\titlerm \scriptfont0=\titlerms \scriptscriptfont0=\titlermss
\textfont1=\titlei \scriptfont1=\titleis \scriptscriptfont1=\titleiss
\textfont2=\titlesy \scriptfont2=\titlesys \scriptscriptfont2=\titlesyss
\textfont\itfam=\titleit \def\it{\fam\itfam\titleit}\rm}
 \ifx\answ\bigans\else scaled\magstep1\fi
\ifx\answ\bigans\def\abstractfont{\tenpoint}\else
\font\abssl=cmsl10 scaled \magstep1
\font\absrm=cmr10 scaled\magstep1 \font\absrms=cmr7 scaled\magstep1
\font\absrmss=cmr5 scaled\magstep1 \font\absi=cmmi10 scaled\magstep1
\font\absis=cmmi7 scaled\magstep1 \font\absiss=cmmi5 scaled\magstep1
\font\abssy=cmsy10 scaled\magstep1 \font\abssys=cmsy7 scaled\magstep1
\font\abssyss=cmsy5 scaled\magstep1 \font\absbf=cmbx10 scaled\magstep1
\skewchar\absi='177 \skewchar\absis='177 \skewchar\absiss='177
\skewchar\abssy='60 \skewchar\abssys='60 \skewchar\abssyss='60
\def\abstractfont{\def\rm{\fam0\absrm}
\textfont0=\absrm \scriptfont0=\absrms \scriptscriptfont0=\absrmss
\textfont1=\absi \scriptfont1=\absis \scriptscriptfont1=\absiss
\textfont2=\abssy \scriptfont2=\abssys \scriptscriptfont2=\abssyss
\textfont\itfam=\bigit \def\it{\fam\itfam\bigit}\def\footnotefont{\tenpoint}%
\textfont\slfam=\abssl \def\sl{\fam\slfam\abssl}%
\textfont\bffam=\absbf \def\bf{\fam\bffam\absbf}\rm}\fi
\def\tenpoint{\def\rm{\fam0\tenrm}
\textfont0=\tenrm \scriptfont0=\sevenrm \scriptscriptfont0=\fiverm
\textfont1=\teni  \scriptfont1=\seveni  \scriptscriptfont1=\fivei
\textfont2=\tensy \scriptfont2=\sevensy \scriptscriptfont2=\fivesy
\textfont\itfam=\tenit
\def\it{\fam\itfam\tenit}\def\footnotefont{\ninepoint}%
\textfont\bffam=\tenbf \def\bf{\fam\bffam\tenbf}\def\sl{\fam\slfam\tensl}\rm}
\font\ninerm=cmr9 \font\sixrm=cmr6 \font\ninei=cmmi9 \font\sixi=cmmi6
\font\ninesy=cmsy9 \font\sixsy=cmsy6 \font\ninebf=cmbx9
\font\nineit=cmti9 \font\ninesl=cmsl9 \skewchar\ninei='177
\skewchar\sixi='177 \skewchar\ninesy='60 \skewchar\sixsy='60
\def\ninepoint{\def\rm{\fam0\ninerm}
\textfont0=\ninerm \scriptfont0=\sixrm \scriptscriptfont0=\fiverm
\textfont1=\ninei \scriptfont1=\sixi \scriptscriptfont1=\fivei
\textfont2=\ninesy \scriptfont2=\sixsy \scriptscriptfont2=\fivesy
\textfont\itfam=\ninei \def\it{\fam\itfam\nineit}\def\sl{\fam\slfam\ninesl}%
\textfont\bffam=\ninebf \def\bf{\fam\bffam\ninebf}\rm}
%
%

\hyphenation{anom-aly anom-alies coun-ter-term coun-ter-terms}
\def\inv{^{\raise.15ex\hbox{${\scriptscriptstyle -}$}\kern-.05em 1}}

\def\Dsl{\,\raise.15ex\hbox{/}\mkern-13.5mu D} 
\def\dsl{\raise.15ex\hbox{/}\kern-.57em\partial}

 \def\Tr{{\rm Tr}}
\font\bigit=cmti10 scaled \magstep1
\def\lspace{\ifx\answ\bigans{}\else\qquad\fi}
\def\lbspace{\ifx\answ\bigans{}\else\hskip-.2in\fi} 
\def\boxeqn#1{\vcenter{\vbox{\hrule\hbox{\vrule\kern3pt\vbox{\kern3pt
           \hbox{${\displaystyle #1}$}\kern3pt}\kern3pt\vrule}\hrule}}}
\def\mbox#1#2{\vcenter{\hrule \hbox{\vrule height#2in
               \kern#1in \vrule} \hrule}}  
%

\def\e#1{{\rm e}^{^{\textstyle#1}}}

\def\darr#1{\raise1.5ex\hbox{$\leftrightarrow$}\mkern-16.5mu #1}

\def\roughly#1{\raise.3ex\hbox{$#1$\kern-.75em\lower1ex\hbox{$\sim$}}}



\def\IB{\relax\hbox{$\inbar\kern-.3em{\rm B}$}}
\def\IC{\relax\hbox{$\inbar\kern-.3em{\rm C}$}}
\def\ID{\relax\hbox{$\inbar\kern-.3em{\rm D}$}}
\def\IE{\relax\hbox{$\inbar\kern-.3em{\rm E}$}}
\def\IF{\relax\hbox{$\inbar\kern-.3em{\rm F}$}}
\def\IG{\relax\hbox{$\inbar\kern-.3em{\rm G}$}}
\def\IGa{\relax\hbox{${\rm I}\kern-.18em\Gamma$}}
\def\IH{\relax{\rm I\kern-.18em H}}
\def\IK{\relax{\rm I\kern-.18em K}}
\def\II{\relax{\rm I\kern-.18em I}}
\def\IL{\relax{\rm I\kern-.18em L}}
\def\IP{\relax{\rm I\kern-.18em P}}
\def\IR{\relax{\rm I\kern-.18em R}}
\def\IZ{\relax\ifmmode\mathchoice {\hbox{\cmss Z\kern-.4em Z}}{\hbox{\cmss
Z\kern-.4em Z}} {\lower.9pt\hbox{\cmsss Z\kern-.4em Z}}
{\lower1.2pt\hbox{\cmsss Z\kern-.4em Z}}\else{\cmss Z\kern-.4em Z}\fi}

\def\IB{\relax{\rm I\kern-.18em B}}
\def\IC{{\relax\hbox{$\inbar\kern-.3em{\rm C}$}}}
\def\ID{\relax{\rm I\kern-.18em D}}
\def\IE{\relax{\rm I\kern-.18em E}}
\def\IF{\relax{\rm I\kern-.18em F}}


\def\CW {{\cal W}}

\def\p{\partial}
\def\pa{\partial}
\def\pb{{\bar{\partial}}}




\def\Tr{{\rm Tr}}


\def\demi{{1\over 2}}


\def\f{\phi}

\def\a{\alpha}
\def\b{\beta}
\def\g{\gamma}  \def\G{\Gamma}
\def\d{\delta}  
\def\m{\mu}

\def\l{\lambda} 

\def\e{\epsilon}

\def\|{\Big|}
\def\({\Big(}   \def\){\Big)}
\def\[{\Big[}   \def\]{\Big]}



\def\paper#1#2#3#4{#1, {\sl #2}, #3 {\tt #4}}

\def\hh{hep-th/}


\def\PLB#1#2#3{Phys. Lett.~{\bf B#1} (#2) #3}
\def\NPB#1#2#3{Nucl. Phys.~{\bf B#1} (#2) #3}
\def\PRL#1#2#3{Phys. Rev. Lett.~{\bf #1} (#2) #3}
\def\CMP#1#2#3{Comm. Math. Phys.~{\bf #1} (#2) #3}
\def\PRD#1#2#3{Phys. Rev.~{\bf D#1} (#2) #3}
\def\MPL#1#2#3{Mod. Phys. Lett.~{\bf #1} (#2) #3}
\def\IJMP#1#2#3{Int. Jour. Mod. Phys.~{\bf #1} (#2) #3}


\def\unlockat{\catcode`\@=11}
\def\lockat{\catcode`\@=12}

\unlockat


\def\newsec#1{\global\advance\secno by1\message{(\the\secno. #1)}
\global\subsecno=0\global\subsubsecno=0\eqnres@t\noindent {\bf\the\secno. #1}
\writetoca{{\secsym} {#1}}\par\nobreak\medskip\nobreak}
\global\newcount\subsecno \global\subsecno=0
\def\subsec#1{\global\advance\subsecno by1\message{(\secsym\the\subsecno.
#1)}
\ifnum\lastpenalty>9000\else\bigbreak\fi\global\subsubsecno=0
\noindent{\it\secsym\the\subsecno. #1}
\writetoca{\string\quad {\secsym\the\subsecno.} {#1}}
\par\nobreak\medskip\nobreak}
\global\newcount\subsubsecno \global\subsubsecno=0
\def\subsubsec#1{\global\advance\subsubsecno by1
\message{(\secsym\the\subsecno.\the\subsubsecno. #1)}
\ifnum\lastpenalty>9000\else\bigbreak\fi
\noindent\quad{\secsym\the\subsecno.\the\subsubsecno.}{#1}
\writetoca{\string\qquad{\secsym\the\subsecno.\the\subsubsecno.}{#1}}
\par\nobreak\medskip\nobreak}

\def\subsubseclab#1{\DefWarn#1\xdef #1{\noexpand\hyperref{}{subsubsection}%
{\secsym\the\subsecno.\the\subsubsecno}%
{\secsym\the\subsecno.\the\subsubsecno}}%
\writedef{#1\leftbracket#1}\wrlabeL{#1=#1}}
\lockat

\def\dbend{\lower3.5pt\hbox{\manual\char127}}


\def\boxit#1{\vbox{\hrule\hbox{\vrule\kern8pt
\vbox{\hbox{\kern8pt}\hbox{\vbox{#1}}\hbox{\kern8pt}}
\kern8pt\vrule}\hrule}}

\def\mathboxit#1{\vbox{\hrule\hbox{\vrule\kern8pt\vbox{\kern8pt
\hbox{$\displaystyle #1$}\kern8pt}\kern8pt\vrule}\hrule}}


\def\inbar{\,\vrule height1.5ex width.4pt depth0pt}

\font\cmss=cmss10 \font\cmsss=cmss10 at 7pt


\lref\simons{ J. Cheeger and J. Simons, {\it Differential Characters and
Geometric Invariants},  Stony Brook Preprint, (1973), unpublished.}

\lref\cargese{ L.~Baulieu, {\it Algebraic quantization of gauge theories},
Perspectives in fields and particles, Plenum Press, eds. Basdevant-Levy,
Cargese Lectures 1983}

\lref\antifields{ L. Baulieu, M. Bellon, S. Ouvry, C.Wallet, Phys.Letters
B252 (1990) 387; M.  Bocchichio, Phys. Lett. B187 (1987) 322;  Phys. Lett. B
192 (1987) 31; R.  Thorn    Nucl. Phys.   B257 (1987) 61. }

\lref\thompson{ George Thompson,  Annals Phys. 205 (1991) 130; J.M.F.
Labastida, M. Pernici, Phys. Lett. 212B  (1988) 56; D. Birmingham, M.Blau,
M. Rakowski and G.Thompson, Phys. Rept. 209 (1991) 129.}

\lref\tonin{ Tonin}

\lref\wittensix{ E.  Witten, {\it New  Gauge  Theories In Six Dimensions},
\hh{9710065}. }

\lref\orlando{ O. Alvarez, L. A. Ferreira and J. Sanchez Guillen, {\it  A New
Approach to Integrable Theories in any Dimension}, hep-th/9710147.}

\lref\wittentopo{ E.  Witten,  {\it  Topological Quantum Field Theory},
\hh9403195, Commun.  Math. Phys.  {117} (1988)353.  }

\lref\wittentwist{ E.  Witten, {\it Supersymmetric Yang--Mills theory on a
four-manifold}, J.  Math.  Phys.  {35} (1994) 5101.}

\lref\west{ L.~Baulieu, P.~West, {\it Six Dimensional TQFTs and  Self-dual
Two-Forms,} Phys.Lett. B {\bf 436 } (1998) 97, /hep-th/9805200}

\lref\bv{ I.A. Batalin and V.A. Vilkowisky,    Phys. Rev.   D28  (1983)
2567\semi M. Henneaux,  Phys. Rep.  126   (1985) 1\semi M. Henneaux and C.
Teitelboim, {\it Quantization of Gauge Systems}
  Princeton University Press,  Princeton (1992).}

\lref\kyoto{ L. Baulieu, E. Bergschoeff and E. Sezgin, Nucl. Phys.
B307(1988)348\semi L. Baulieu,   {\it Field Antifield Duality, p-Form Gauge
Fields
   and Topological Quantum Field Theories}, hep-th/9512026,
   Nucl. Phys. B478 (1996) 431.  }

\lref\sourlas{ G. Parisi and N. Sourlas, {\it Random Magnetic Fields,
Supersymmetry and Negative Dimensions}, Phys. Rev. Lett.  43 (1979) 744;
Nucl.  Phys.  B206 (1982) 321.  }

\lref\SalamSezgin{ A.  Salam  and  E.  Sezgin, {\it Supergravities in
diverse dimensions}, vol.  1, p. 119\semi P.  Howe, G.  Sierra and P.
Townsend, Nucl Phys B221 (1983) 331.}

\lref\nekrasov{ A. Losev, G. Moore, N. Nekrasov, S. Shatashvili, {\it
Four-Dimensional Avatars of Two-Dimensional RCFT},  hep-th/9509151, Nucl.
Phys.  Proc.  Suppl.   46 (1996) 130\semi L.  Baulieu, A.  Losev,
N.~Nekrasov  {\it Chern-Simons and Twisted Supersymmetry in Higher
Dimensions},  hep-th/9707174, to appear in Nucl.  Phys.  B.  }

\lref\WitDonagi{R.~ Donagi, E.~ Witten, ``Supersymmetric Yang--Mills Theory
and Integrable Systems'', hep-th/9510101, Nucl. Phys.{\bf B}460 (1996)
299-334}
\lref\Witfeb{E.~ Witten, ``Supersymmetric Yang--Mills Theory On A
Four-Manifold,''  hep-th/9403195; J. Math. Phys. {\bf 35} (1994) 5101.}
\lref\Witgrav{E.~ Witten, ``Topological Gravity'', Phys.Lett.206B:601, 1988}
\lref\witaffl{I. ~ Affleck, J.A.~ Harvey and E.~ Witten,
        ``Instantons and (Super)Symmetry Breaking
        in $2+1$ Dimensions'', Nucl. Phys. {\bf B}206 (1982) 413}
\lref\wittabl{E.~ Witten,  ``On $S$-Duality in Abelian Gauge Theory,''
hep-th/9505186; Selecta Mathematica {\bf 1} (1995) 383}
\lref\wittgr{E.~ Witten, ``The Verlinde Algebra And The Cohomology Of The
Grassmannian'',  hep-th/9312104}
\lref\wittenwzw{E. Witten, ``Non abelian bosonization in two dimensions,''
Commun. Math. Phys. {\bf 92} (1984)455 }
\lref\witgrsm{E. Witten, ``Quantum field theory, grassmannians and algebraic
curves,'' Commun.Math.Phys.113:529,1988}
\lref\wittjones{E. Witten, ``Quantum field theory and the Jones
polynomial,'' Commun.  Math. Phys., 121 (1989) 351. }
\lref\witttft{E.~ Witten, ``Topological Quantum Field Theory", Commun. Math.
Phys. {\bf 117} (1988) 353.}
\lref\wittmon{E.~ Witten, ``Monopoles and Four-Manifolds'', hep-th/9411102}
\lref\Witdgt{ E.~ Witten, ``On Quantum gauge theories in two dimensions,''
Commun. Math. Phys. {\bf  141}  (1991) 153}
\lref\witrevis{E.~ Witten,
 ``Two-dimensional gauge theories revisited'', hep-th/9204083; J. Geom.
Phys. 9 (1992) 303-368}
\lref\Witgenus{E.~ Witten, ``Elliptic Genera and Quantum Field Theory'',
Comm. Math. Phys. 109(1987) 525. }
\lref\OldZT{E. Witten, ``New Issues in Manifolds of SU(3) Holonomy,'' {\it
Nucl. Phys.} {\bf B268} (1986) 79 \semi J. Distler and B. Greene, ``Aspects
of (2,0) String Compactifications,'' {\it Nucl. Phys.} {\bf B304} (1988) 1
\semi B. Greene, ``Superconformal Compactifications in Weighted Projective
Space,'' {\it Comm. Math. Phys.} {\bf 130} (1990) 335.}
\lref\bagger{E.~ Witten and J. Bagger, Phys. Lett. {\bf 115B}(1982) 202}
\lref\witcurrent{E.~ Witten,``Global Aspects of Current Algebra'',
Nucl.Phys.B223 (1983) 422\semi ``Current Algebra, Baryons and Quark
Confinement'', Nucl.Phys. B223 (1993) 433}
\lref\Wittreiman{S.B. Treiman, E. Witten, R. Jackiw, B. Zumino, ``Current
Algebra and Anomalies'', Singapore, Singapore: World Scientific ( 1985) }
\lref\Witgravanom{L. Alvarez-Gaume, E.~ Witten, ``Gravitational Anomalies'',
Nucl.Phys.B234:269,1984. }

\lref\nicolai{\paper {H.~Nicolai}{New Linear Systems for 2D Poincar\'e
Supergravities}{\NPB{414}{1994}{299},}{\hh 9309052}.}



\lref\baex{\paper {L.~Baulieu, B.~Grossman}{Monopoles and Topological Field
Theory}{\PLB{214}{1988}{223}.}{}\paper {L.~Baulieu}{Chern-Simons
Three-Dimensional and
Yang--Mills-Higgs Two-Dimensional Systems as Four-Dimensional Topological
Quantum Field Theories}{\PLB{232}{1989}{473}.}{}}

\lref\bg{\paper {L.~Baulieu, B.~Grossman}{Monopoles and Topological Field
Theory}{\PLB{214}{1988}{223}.}{}}

\lref\seibergsix{\paper {N.~Seiberg}{Non-trivial Fixed Points of The
Renormalization Group in Six
 Dimensions}{\PLB{390}{1997}{169}}{\hh 9609161}\semi
\paper {O.J.~Ganor, D.R.~Morrison, N.~Seiberg}{
  Branes, Calabi-Yau Spaces, and Toroidal Compactification of the N=1
  Six-Dimensional $E_8$ Theory}{\NPB{487}{1997}{93}}{\hh 9610251}\semi
\paper {O.~Aharony, M.~Berkooz, N.~Seiberg}{Light-Cone
  Description of (2,0) Superconformal Theories in Six
  Dimensions}{Adv. Theor. Math. Phys. {\bf 2} (1998) 119}{\hh 9712117.}}

\lref\cs{\paper {L.~Baulieu}{Chern-Simons Three-Dimensional and
Yang--Mills-Higgs Two-Dimensional Systems as Four-Dimensional Topological
Quantum Field Theories}{\PLB{232}{1989}{473}.}{}}

\lref\beltrami{\paper {L.~Baulieu, M.~Bellon}{Beltrami Parametrization and
String Theory}{\PLB{196}{1987}{142}}{}\semi
\paper {L.~Baulieu, M.~Bellon, R.~Grimm}{Beltrami Parametrization For
Superstrings}{\PLB{198}{1987}{343}}{}\semi
\paper {R.~Grimm}{Left-Right Decomposition of Two-Dimensional Superspace
Geometry and Its BRS Structure}{Annals Phys. {\bf 200} (1990) 49.}{}}

\lref\bbg{\paper {L.~Baulieu, M.~Bellon, R.~Grimm}{Left-Right Asymmetric
Conformal Anomalies}{\PLB{228}{1989}{325}.}{}}

\lref\bonora{\paper {G.~Bonelli, L.~Bonora, F.~Nesti}{String Interactions
from Matrix String Theory}{\NPB{538}{1999}{100},}{\hh 9807232}\semi
\paper {G.~Bonelli, L.~Bonora, F.~Nesti, A.~Tomasiello}{Matrix String Theory
and its Moduli Space}{}{\hh 9901093.}}

\lref\corrigan{\paper {E.~Corrigan, C.~Devchand, D.B.~Fairlie,
J.~Nuyts}{First Order Equations for Gauge Fields in Spaces of Dimension
Greater Than Four}{\NPB{214}{452}{1983}.}{}}

\lref\acha{\paper {B.S.~Acharya, M.~O'Loughlin, B.~Spence}{Higher
Dimensional Analogues of Donaldson-Witten Theory}{\NPB{503}{1997}{657},}{\hh
9705138}\semi
\paper {B.S.~Acharya, J.M.~Figueroa-O'Farrill, M.~O'Loughlin,
B.~Spence}{Euclidean
  D-branes and Higher-Dimensional Gauge   Theory}{\NPB{514}{1998}{583},}{\hh
  9707118.}}

\lref\Witr{\paper{E.~Witten}{Introduction to Cohomological Field   Theories}
{Lectures at Workshop on Topological Methods in Physics (Trieste, Italy, Jun
11-25, 1990), \IJMP{A6}{1991}{2775}.}{}}

\lref\ohta{\paper {L.~Baulieu, N.~Ohta}{Worldsheets with Extended
Supersymmetry} {\PLB{391}{1997}{295},}{\hh 9609207}.}

\lref\gravity{\paper {L.~Baulieu}{Transmutation of Pure 2-D Supergravity
Into Topological 2-D Gravity and Other Conformal Theories}
{\PLB{288}{1992}{59},}{\hh 9206019.}}

\lref\wgravity{\paper {L.~Baulieu, M.~Bellon, R.~Grimm}{Some Remarks on  the
Gauging of the Virasoro and   $w_{1+\infty}$
Algebras}{\PLB{260}{1991}{63}.}{}}

\lref\fourd{\paper {E.~Witten}{Topological Quantum Field
Theory}{\CMP{117}{1988}{353}}{}\semi
\paper {L.~Baulieu, I.M.~Singer}{Topological Yang--Mills Symmetry}{Nucl.
Phys. Proc. Suppl. {\bf 15B} (1988) 12.}{}}

\lref\topo{\paper {L.~Baulieu}{On the Symmetries of Topological Quantum Field
Theories}{\IJMP{A10}{1995}{4483},}{\hh 9504015}\semi
\paper {R.~Dijkgraaf, G.~Moore}{Balanced Topological Field
Theories}{\CMP{185}{1997}{411},}{\hh 9608169.}}

\lref\wwgravity{\paper {I.~Bakas} {The Large $N$ Limit   of Extended
Conformal Symmetries}{\PLB{228}{1989}{57}.}{}}

\lref\wwwgravity{\paper {C.M.~Hull}{Lectures on $\CW$-Gravity,
$\CW$-Geometry and
$\CW$-Strings}{}{\hh 9302110}, and~references therein.}

\lref\wvgravity{\paper {A.~Bilal, V.~Fock, I.~Kogan}{On the origin of
$W$-algebras}{\NPB{359}{1991}{635}.}{}}

\lref\surprises{\paper {E.~Witten} {Surprises with Topological Field
Theories} {Lectures given at ``Strings 90'', Texas A\&M, 1990,}{Preprint
IASSNS-HEP-90/37.}}

\lref\stringsone{\paper {L.~Baulieu, M.B.~Green, E.~Rabinovici}{A Unifying
Topological Action for Heterotic and  Type II Superstring  Theories}
{\PLB{386}{1996}{91},}{\hh 9606080.}}

\lref\stringsN{\paper {L.~Baulieu, M.B.~Green, E.~Rabinovici}{Superstrings
from   Theories with $N>1$ World Sheet Supersymmetry}
{\NPB{498}{1997}{119},}{\hh 9611136.}}

\lref\bks{\paper {L.~Baulieu, H.~Kanno, I.~Singer}{Special Quantum Field
Theories in Eight and Other Dimensions}{\CMP{194}{1998}{149},}{\hh
9704167}\semi
\paper {L.~Baulieu, H.~Kanno, I.~Singer}{Cohomological Yang--Mills Theory
  in Eight Dimensions}{ Talk given at APCTP Winter School on Dualities in
String Theory (Sokcho, Korea, February 24-28, 1997),} {\hh 9705127.}}

\lref\witdyn{\paper {P.~Townsend}{The eleven dimensional supermembrane
revisited}{\PLB{350}{1995}{184},}{\hh9501068}\semi
\paper{E.~Witten}{String Theory Dynamics in Various Dimensions}
{\NPB{443}{1995}{85},}{\hh 9503124}.}

\lref\bfss{\paper {T.~Banks, W.Fischler, S.H.~Shenker,
L.~Susskind}{$M$-Theory as a Matrix Model~:
A~Conjecture}{\PRD{55}{1997}{5112},} {\hh9610043.}}

\lref\seiberg{\paper {N.~Seiberg}{Why is the Matrix Model
Correct?}{\PRL{79}{1997}{3577},} {\hh 9710009.}}

\lref\sen{\paper {A.~Sen}{$D0$ Branes on $T^n$ and Matrix Theory}{Adv.
Theor. Math. Phys.~{\bf 2} (1998) 51,} {\hh 9709220.}}

\lref\laroche{\paper {L.~Baulieu, C.~Laroche} {On Generalized Self-Duality
Equations Towards Supersymmetric   Quantum Field Theories Of
Forms}{\MPL{A13}{1998}{1115},}{\hh  9801014.}}

\lref\bsv{\paper {M.~Bershadsky, V.~Sadov, C.~Vafa} {$D$-Branes and
Topological Field Theories}{\NPB{463} {1996}{420},}{\hh9511222.}}

\lref\vafapuzz{\paper {C.~Vafa}{Puzzles at Large N}{}{\hh 9804172.}}

\lref\dvv{\paper {R.~Dijkgraaf, E.~Verlinde, H.~Verlinde} {Matrix String
Theory}{\NPB{500}{1997}{43},} {\hh9703030.}}

\lref\wynter{\paper {T.~Wynter}{Gauge Fields and Interactions in Matrix
String Theory}{\PLB{415}{1997}{349},}{\hh9709029.}}

\lref\kvh{\paper {I.~Kostov, P.~Vanhove}{Matrix String Partition
Functions}{}{\hh9809130.}}

\lref\ikkt{\paper {N.~Ishibashi, H.~Kawai, Y.~Kitazawa, A.~Tsuchiya} {A
Large $N$ Reduced Model as Superstring}{\NPB{498} {1997}{467},}{\hh
9612115.}}

\lref\ss{\paper {S.~Sethi, M.~Stern} {$D$-Brane Bound States
Redux}{\CMP{194}{1998} {675},}{\hh 9705046.}}

\lref\mns{\paper {G.~Moore, N.~Nekrasov, S.~Shatashvili} {$D$-particle Bound
States and Generalized Instantons}{} {\hh 9803265.}}

\lref\bsh{\paper {L.~Baulieu, S.~Shatashvili} {Duality from Topological
Symmetry}{} {\hh 9811198.}}

\lref\pawu{ {G.~Parisi, Y.S.~Wu} {}{ Sci. Sinica  {\bf 24} {(1981)} {484}.}}

\lref\coulomb{ {L.~Baulieu, D.~Zwanziger, }   {\it Renormalizable Non-Covariant
Gauges and Coulomb Gauge Limit}, {Nucl.Phys. B {\bf 548 } (1999) 527.} {\hh
9807024}.}

\lref\rcoulomb{ {D.~Zwanziger, }   {\it Renormalization in the Coulomb
gauge and order parameter for confinement in QCD}, {Nucl.Phys. B {\bf 538
} (1998) 237.} {}}

\lref\horne{ {J.H.~Horne, }   {\it
Superspace versions of Topological Theories}, {Nucl.Phys. B {\bf 318
} (1989) 22.} {}}

\lref\sto{ {S.~Ouvry, R.~Stora, P.~Van~Baal }   {\it
}, {Phys. Lett. B {\bf 220
} (1989) 159;} {}{ R.~Stora, {\it Exercises in   Equivariant Cohomology},
In Quabtum Fields and Quantum Space Time, Edited
by 't Hooft et al., Plenum Press, New York, 1997}            }

\lref\dzvan{ {D.~Zwanziger, }   {\it Vanishing of zero-momentum lattice
gluon propagator and color confinement}, {Nucl.Phys. B {\bf 364 }
(1991) 127.} }

\lref\dan{ {D.~Zwanziger},  {\it Covariant Quantization of Gauge
Fields without Gribov Ambiguity}, {Nucl. Phys. B {\bf   192}, (1981)
{259}.}{}}

\lref\danzinn{  {J.~Zinn-Justin, D.~Zwanziger, } {}{Nucl. Phys. B  {\bf
295} (1988) {297}.}{}}

\lref\danlau{ {L.~Baulieu, D.~Zwanziger, } {\it Equivalence of Stochastic
Quantization and the-Popov Ansatz,
  }{Nucl. Phys. B  {\bf 193 } (1981) {163}.}{}}

\lref\munoz{ { A.~Munoz Sudupe, R. F. Alvarez-Estrada, } {}
Phys. Lett. {\bf 164} (1985) 102; {} {\bf 166B} (1986) 186. }

\lref\okano{ { K.~Okano, } {}
Nucl. Phys. {\bf B289} (1987) 109; {} Prog. Theor. Phys.
suppl. {\bf 111} (1993) 203. }

\lref\singer{
 I.M. Singer, { Comm. of Math. Phys. {\bf 60} (1978) 7.}}

\lref\neu{ {H.~Neuberger,} {Phys. Lett. B {\bf 295}
(1987) {337}.}{}}

\lref\testa{ {M.~Testa,} {}{Phys. Lett. B {\bf 429}
(1998) {349}.}{}}

\lref\Martin{ L.~Baulieu and M. Schaden, {\it Gauge Group TQFT and Improved
Perturbative Yang--Mills Theory}, {  Int. Jour. Mod.  Phys. A {\bf  13}
(1998) 985},   hep-th/9601039.}

\lref\baugros{ {L.~Baulieu, B.~Grossman, } {\it A topological Interpretation
of  Stochastic Quantization} {Phys. Lett. B {\bf  212} {(1988)} {351}.}}

\lref\bautop{ {L.~Baulieu}{ \it Stochastic and Topological Field Theories},
{Phys. Lett. B {\bf   232} (1989) {479}}{}; {}{ \it Topological Field Theories
And Gauge Invariance in Stochastic Quantization}, {Int. Jour. Mod.  Phys. A
{\bf  6} (1991) {2793}.}{}}

\lref\bautopr{  {L.~Baulieu, B.~Grossman, } {\it A topological Interpretation
of  Stochastic Quantization} {Phys. Lett. B {\bf  212} {(1988)} {351}};
 {L.~Baulieu}{ \it Stochastic and Topological Field Theories},
{Phys. Lett. B {\bf   232} (1989) {479}}{}; {}{ \it Topological Field Theories
And Gauge Invariance in Stochastic Quantization}, {Int. Jour. Mod.  Phys. A
{\bf  6} (1991) {2793}.}{}}

\lref\samson{ {L.~Baulieu, S.L.~Shatashvili, { \it Duality from Topological
Symmetry}, {JHEP {\bf 9903} (1999) 011, hep-th/9811198.}}}{}

\lref\halpern{ {H.S.~Chan, M.B.~Halpern}{}, {Phys. Rev. D {\bf   33} (1986)
{540}.}}

\lref\yue{ {Yue-Yu}, {Phys. Rev. D {\bf   33} (1989) {540}.}}

\lref\neuberger{ {H.~Neuberger,} {\it Non-perturbative gauge Invariance},
{ Phys. Lett. B {\bf 175} (1986) {69}.}{}}

\lref\gribov{  {V.N.~Gribov,} {}{Nucl. Phys. B {\bf 139} (1978) {1}.}{}}

\lref\huffel{ {P.H.~Daamgard, H. Huffel},  {}{Phys. Rep. {\bf 152} (1987)
{227}.}{}}

\lref\creutz{ {M.~Creutz},  {\it Quarks, Gluons and  Lattices, }  Cambridge
University Press 1983, pp 101-107.}

\lref\zinn{ {J. ~Zinn-Justin, }  {Nucl. Phys. B {\bf  275} (1986) {135}.}}

\lref\gozzi{ {E. ~Gozzi,} {\it Functional Integral approach to Parisi--Wu
Quantization: Scalar Theory,} { Phys. Rev. {\bf D28} (1983) {1922}.}}

\lref\shamir{  {Y.~Shamir,  } {\it Lattice Chiral Fermions
  }{ Nucl.  Phys.  Proc.  Suppl.  {\bf } 47 (1996) 212,  hep-lat/9509023;
V.~Furman, Y.~Shamir, Nucl.Phys. B {\bf 439 } (1995), hep-lat/9405004.}}

 \lref\kaplan{ {D.B.~Kaplan, }  {\it A Method for Simulating Chiral
Fermions on the Lattice,} Phys. Lett. B {\bf 288} (1992) 342; {\it Chiral
Fermions on the Lattice,}  {  Nucl. Phys. B, Proc. Suppl.  {\bf 30} (1993)
597.}}

\lref\neubergerr{ {H.~Neuberger, } {\it Chirality on the Lattice},
hep-lat/9808036.}

\lref\neubergers{ {Rajamani Narayanan, Herbert Neuberger,} {\it INFINITELY MANY
    REGULATOR FIELDS FOR CHIRAL FERMIONS.}
    Phys.Lett.B302:62-69,1993.
    [HEP-LAT 9212019]}

\lref\neubergert{ {Rajamani Narayanan, Herbert Neuberger,}{\it CHIRAL FERMIONS
    ON THE LATTICE.}
    Phys.Rev.Lett.71:3251-3254,1993.
    [HEP-LAT 9308011]}

\lref\neubergeru{ {Rajamani Narayanan, Herbert Neuberger,}{\it A CONSTRUCTION
OF LATTICE CHIRAL GAUGE THEORIES.}
    Nucl.Phys.B443:305-385,1995.
    [HEP-TH 9411108]}

\lref\neubergerv{ {Herbert Neuberger,}{\it EXACTLY MASSLESS QUARKS ON THE
    LATTICE.}
    Phys. Lett. B417 (1998) 141-144.
    [HEP-LAT 9707022]}

%

\lref\neubergerw{ {Herbert Neuberger,}{\it CHIRAL FERMIONS ON THE
LATTICE.}
    Nucl. Phys. B, Proc. Suppl. 83-84 (2000) 67-76.
    [HEP-LAT 9909042]}

\lref\zbgr {L.~Baulieu and D. Zwanziger, {\it QCD$_4$ From a
Five-Dimensional Point of View},    hep-th/9909006.}
\lref\bgz {P. A. Grassi, L.~Baulieu and D. Zwanziger, {\it Gauge and
Topological
Symmetries in the Bulk Quantization of Gauge Theories},    hep-th/0006036.}


\vskip -1cm
\Title{
\vbox
{\baselineskip 10pt
\hbox{hep-th/0012103}
\hbox{LPTHE-00-45}
\hbox{RUNHETC-2000-53}
\hbox{NYU-TH-30.5.00}
\vskip -2cm
 \hbox{   }}}
{\vbox{\vskip -30 true pt
\centerline{
   }
\medskip
 \centerline{From stochastic quantization to bulk quantization:  }
\centerline{Schwinger-Dyson equations and  S-matrix }
\vskip -1cm
\medskip
\vskip4pt }}
\centerline{{\bf Laurent Baulieu}$^{\star     }$
  and  {\bf  Daniel
Zwanziger}$^{ \dag}$}
\centerline{baulieu@lpthe.jussieu.fr, Daniel.Zwanziger@nyu.edu}
\vskip 0.1cm
\centerline{\it $^{\star}$LPTHE, Universit{\'e}s P. \& M. Curie (Paris~VI) et
D. Diderot (Paris~VII), Paris,  France}
\centerline{\it
 $^{\star}$
Department of Physics and Astronomy, Rutgers University, Piscataway, NJ
088555-0849, USA }
\centerline{\it $^{\dag}$   Physics Department, New York University,
New-York,  NY 10003,  USA}
\vskip -1cm

\medskip
\vskip  1.6cm
	In stochastic quantization, ordinary 4-dimensional Euclidean quantum
field theory is expressed as a functional integral over fields in 5
dimensions with a fictitious 5th time.  This is advantageous, in particular
for gauge theories, because it allows a different type of gauge fixing that
avoids the Gribov problem.    Traditionally, in this approach, the
fictitious 5th time is the analog of computer time in a Monte Carlo
simulation of 4-dimensional Euclidean fields.  A Euclidean
probability distribution  which depends on the 5th time relaxes to an
equilibrium distribution.  However a broader framework, which we
call``bulk quantization", is required for extension to fermions, and for
the increased  power afforded by the higher symmetry of the
5-dimensional action that is topological when expressed in terms of
auxiliary fields.  Within the broader framework, we give a direct proof by
means of Schwinger-Dyson equations that a time-slice of the
5-dimensional theory is equivalent to the usual 4-dimensional theory.
The proof does not rely on the conjecture that the relevant stochastic
process relaxes to an equilibrium distribution. Rather, it depends on the
higher symmetry of the 5-dimensional action which includes a
BRST-type topological invariance, and invariance under translation and
inversion in the 5-th time. We express the physical S-matrix directly in
terms of the truncated 5-dimensional correlation functions, for which
``going off the mass-shell'' means going from the 3 physical degrees of
freedom to 5 independent variables. We derive the Landau-Cutokosky
rules of the 5-dimensional theory which include the physical unitarity
relation.

\Date{\ }

\def\e{\epsilon}
\def\demi{{1\over 2}}
\def\quart{{1\over 4}}
\def\pa{\partial}
\def\a{\alpha}
\def\b{\beta}
\def\d{\delta}

\def\m{\mu}

\def\l{\lambda}

\def\pb{\bar{\psi}}

\def\t{\theta}

\newsec{Introduction}

	There are a number of cases where it is helpful to increase
the number of dimensions above what seems to be required.  An example
is the use of complex numbers, with an ``imaginary'' component, to
solve a problem involving real numbers. There is also Feynman's
expression of three-dimensional quantum mechanical perturbation theory
by four-dimensional space-time integrals.  For the quantization of
gauge fields, it is useful to repeat Feynman's idea and express
ordinary ``four-dimensional'' quantum field theory in terms of a
functional integral in five dimensions, which allows one to overcome
the Gribov problem \pawu, \dan.  Use of the 5th dimension in quantum
field theory began with the paper of Parisi and Wu~\pawu\ which
introduced a method that has become known as stochastic quantization.
Here one regards the 4-dimensional Euclidean probability distribution
$N\exp[ - S(\f)]$ as the equilibrium Boltzmann distribution to which a
stochastic process relaxes.  Such a process is described by the
Fokker-Planck equation:
\eqn\fopl{\eqalign{
\dot{P} =  { {\p} \over {\p \f_i} } \Big[
\Big({ {\p} \over {\p \f_i} } + { {\p S} \over {\p \f_i} } \Big) P \Big] ,
}}
where $P = P(\f, t)$ is a probability that evolves in a ``fictional'' or
``fifth'' time $t$.  In the notation used here,
the discrete index $i$ represents the usual 4-dimensional Euclidean
space-time variable $x_\m$, with $\m = 1,...4$, and possibly other
labels, and for the field variable
we write  $\f_i$ instead of $\f(x)$.  More generally we have
\eqn\foplk{\eqalign{
\dot{P} =  { {\p} \over {\p \f_i} } \Big[ K_{ij}(\f)
\Big({ {\p} \over {\p \f_j} } + { {\p S} \over {\p \f_j} } \Big) P \Big],
}}
where $K_{ij}(\f)$ is an appropriate kernel.  One sees immediately
that the desired Euclidean distribution $N \exp[ - S(\f)]$ is a
time-independent solution of this equation.

	Stochastic quantization relies on the assumption that starting from
any initial normalized probability distribution
$P(\f, 0) = P_0(\f)$, the solution $P(\f, t)$ relaxes to an equilibrium
distribution $P_{\rm eq}(\f)$ which moreover is unique,
\eqn\rlx{\eqalign{
 \lim_{t \to \infty}P(\f, t) = P_{\rm eq}(\f) = N\exp[-S(\f)],
}}
and stochastic quantization is
realized concretely in numerical simulations of quantum field theory by
the Monte--Carlo method, where the 5th time corresponds to the
number of sweeps of the lattice.
According to the Frobenius theorem, relaxation to a unique equilibrium
distribution does hold for a large
class of discrete Markov processes with a finite number of variables.
However it is not known whether the Frobenius theorem applies in
the field-theoretic context with its infinite number of degrees of
freedom.

This approach has been considerably elaborated over the years
\gozzi,\huffel,\zinn,\halpern,\baugros, and new ideas have
emerged.  Just as in conventional quantum field theory one does not deal
directly with the Schroedinger equation but rather with correlation
functions in 4 dimensions calculated from a 4-dimensional functional
integral, likewise in stochastic quantization, it is more efficient to
deal, not
with the Fokker-Planck equation itself, but rather with correlation
functions in 5 dimensions calculated from a 5-dimensional functional
integral with a 5-dimensional local action $I = \int d^4x dt {\cal
L}$.  It is convenient to let the 5th time run from $t = - \infty$ to
$t = + \infty$, and to exploit invariance under time translation and,
significantly, under inversion in the 5th time, $t \to -t$, and there
are no initial conditions to be specified.  (This corresponds to a
stationary process that is at equilibrium for all 5th time.)  It turns
out that for such purposes as renormalization, the most effective
functional integral representation is  as  a topological quantum field
theory (i.~e.~with a BRST-exact action) that necessarily involves
auxiliary and ghost fields.  These fields do not have an immediate
stochastic interpretation.  The stochastic interpretation is also lost
in the extension of the theory to fermi-dirac fields.  Thus the
relation to a stochastic and relaxation process is rather tenuous, and
one feels the need of a more fundamental approach for quantization
with an additional time.

Inspired by the recent developments in holography principles, we
have recently revisited stochastic quantization of gauge theory in the
context of topological quantum field theory \zbgr,\bgz. The beauty of
the construction  suggests to us to further investigate the
guiding principles of quantization with an additional time.  In the
present article we shall give a direct proof by means of
Schwinger-Dyson equations that a time-slice of the 5-dimensional
theory is equivalent to the usual 4-dimensional Euclidean theory.  It
holds for fermi as well as bose fields.  Our proof does {\it not}
involve any relaxation, but relies instead on the higher symmetry of
the 5-dimensional action, including its BRST-type topological
invariance, and invariance under translation in the 5th time.
Somewhat surprisingly, invariance under inversion in the 5th time, $t
\to -t$, plays an essential role in our proof.  We also show that
symmetry under inversion in the 5th time selects the class of
5-dimensional theories that are equivalent on a time-slice to a local
4-dimensional theory.  We shall also express the physical
S-matrix directly in terms of the truncated correlation functions of
the 5-dimensional theory, eq.~(4.34) below.  When momentum vectors of
physical particles are ``taken off the mass shell'', they acquire not
4 but 5 components.  We also elaborate on the Cutkowski--Landau rules
along the same lines, as well as on the conservation laws when there
are global symmetries.

	Because the connection to a stochastic process is not obvious in the
present approach, this name no longer seems appropriate.  We call our
approach ``bulk'' quantization.  Here we regard the 5-dimensional space
with local topological action as the bulk, with the physical 4-space living
on a time-slice.  In the present article, we deal only with a theory of
non-gauge type. In this case there is an exact equivalence of the 4- and
5-dimensional formulations, and we already find quite remarkable that
supersymmetry -- in reality the consequences of the BRST symmetry --
allows one to directly prove the results without having to invoke the
Frobenius theorem.   This opens interesting questions about the
convergence of correlation functions in the limit of equal (fifth)
time. For gauge theories, much more must be done, which we will present
in a separate publication.  Whereas for theories of non-gauge type the
equivalence of the standard and bulk quantization is exact, for gauge
theories bulk quantization allows a different type a gauge fixing that
overcomes the Gribov problem.

\newsec{Equivalence of standard and bulk quantization}

\def\O{{\cal O}}
\def\cD{{\cal D}}

	We consider a theory of a field or set of fields $\f(x)$ in $d$
Euclidean space-time
dimensions, for $x = x_\m$, with $\m = 1,...d$,  defined by a local action
 \eqn\dact{\eqalign{ S[\f] & = \int d^d x \ {\cal L}_d .}}
For example, we may take
 \eqn\dact{\eqalign{	{\cal L}_d & = (1/2) (\p_\m \f)^2 + (1/2) m^2 \f^2
+ (1/4) g \f^4 . }}
Expectation-values are calculated from
\eqn\eval{\eqalign{
 \langle \O \rangle_d \equiv N \int d\f  \ \O \exp(-S)   \ ,
}}
for any observable $\O = \O(\f)$.

This theory is completely described, at least perturbatively,
by the familiar set of Schwinger-Dyson (SD)
equations.  They follow from the identity
\eqn\idtt{\eqalign{
0 =  \int d\f \ { {\p [ \O \exp( - S)]} \over {\p \f_i} }
}}
\eqn\sd{\eqalign{ \langle { {\p \O} \over {\p \f_i} }
    - \O { {\p S} \over {\p \f_i} } \rangle_d = 0
}}
that holds for all observables $\O(\f)$.  Here and below the discrete index
$i$ or $j$ etc.
represents the continuous index $x_\m$, $\m = 1,...d$ and any internal indices.
It is sufficient to take $\O(\f) = \exp(j_i \f_i)$, where the $j_i$ are
arbitrary
sources, to generate a complete set of Schwinger-Dyson equations that
determine the partition function $Z(j) = \langle \exp(j_i \f_i) \rangle$.

	Bulk quantization of the same theory is expressed in terms of a
$(d+1)$-dimensional theory that involves a quartet of fields
$\f(x, t), \psi(x, t), \pb(x, t), b(x, t)$, with ghost number $0, 1, -1, 0$
respectively, the
ghost and anti-ghost fields being fermionic.  A topological BRST operator $s$
acts on these fields according to
\eqn\defs{\eqalign{
s\f & = \psi, \ \ \ s\psi = 0 \cr
s\pb & = b, \ \ \ \ sb = 0,}}
and obviously satisfies $s^2 = 0$.  A $(d+1)$-dimensional action is defined by
\eqn\tact{\eqalign{
I_{\rm tot} & = \int dt \ \demi \dot{S} + I  \cr
	I  & \equiv \int dt \  s \{  \ \pb_i \ [ \  \dot{\f}_i
+ K_{ij}({ {\p S} \over {\p \f_j} } + b_j ) \ ]  \  \} \ ,
}}
where $\dot{S}  \equiv { {\p S} \over { \p t } }$,
 $\dot{\f} \equiv { {\p \f} \over { \p t } }$ etc.  It is topological in
the sense that it is an
exact derivative plus an
$s$-exact term.  In this section we consistently replace the continuous
$d$-dimensional variable $x$ by the discrete index
$i$, but we keep the continuous variable $t$, so the fields are represented by
$\f_i(t), \psi_i(t), \pb_i(t), b_i(t)$.  The action $I$
has the expansion
\eqn\eact{\eqalign{
I  = \int dt \ \{ \ b_i[  \ \dot{\f}_i
+ K_{ij}({ {\p S} \over {\p \f_j} } + b_j ) \ ]
	- \pb_i  \ ( \dot{\psi}_i + L_{il}\psi_l ) \ \}  \ ,
}}
where
\eqn\defl{\eqalign{
L_{il} \equiv K_{ij}{ {\p^2 S} \over {\p \f_j \p \f_l} }
+ { {\p K_{ij}} \over {\p \f_l} }({ {\p S} \over {\p \f_j} } + b_j )  \ .
}}

	In $d$ = 4 dimensions, the kernel $K_{ij}$ is severely restricted
by renormalizability.
For a scalar theory we will have $K = {\rm const}$.  The motive for
introducing the
kernel, apart from generality, is that in a theory with a Dirac spinor $q$
it is helpful for
convergence of Feynman integrals to take
$K = - \g_\m \p_\m +$ (non-derivative), so that the highest derivative of
$q$, contained
in
\eqn\dirspin{\eqalign{
K  { {\d S} \over {\d \bar{q}} }
= (- \g_\m \p_\m + ...)(\g_\m \p_\m + ...)q = (- \p^2 + ...)q
}}
is a positive elliptic operator.

	Contact with the physical theory in $d$ dimensions is made by
requiring that physical observables $\O$ be functions of $\f$ that are
restricted to a time-slice $t = t_0$.  By time-translation invariance
we may take $t_0 = 0$.  Thus the allowed physical observables are of
the form $\O = \O(\f(0))$.  Derivatives with respect to $t$ are not
allowed. Note that the topological invariance under reparametrization of
the the variable $t$ is also an invariance  of the (d+1)-dimensional
action ${{1}\over{2}}\int  dt \dot S$.  We wish to emphasize that
observables are {\it not} of topological type: they are {\it not} $s$-exact.

	We now come to the essential point.  {\it Statement}:  The
$s$-invariant  theory in $(d+1)$ dimensions, with action $I_{\rm tot}$
and observables restricted to a time-slice, is identical to the
$d$-dimensional Euclidean theory with action~$S$, provided only that
the kernel $K_{ij}$ is symmetric and transverse,
\eqn\kernel{\eqalign{  K_{ij} = K_{ji}, \ \ \ \ \ \  {
{\p K_{ij}} \over {\p \f_j} } = 0.  }}
In other words we assert that the two theories
give the same expectation values for all physical observables,
\eqn\eqv{\eqalign{
\langle \O(\f) \rangle_d = \langle \O(\f(0)) \rangle_{d+1} \ ,
}}
where $\langle \O(\f) \rangle_d$ is defined in \eval\ and
\eqn\expc{\eqalign{
\langle \O \rangle_{d+1} \equiv N \int \cD\f  \cD b  \cD \psi  \cD \pb
\ \O \exp I_{\rm tot}
\ . }}
[Note that the weights are $\exp(-S)$ and $\exp(+I_{\rm tot})$.]

	{\it Proof}: It is sufficient to establish that the SD
equations \sd\ are satisfied for
the expectation-values
$\langle \O(\f(0)) \rangle_{d+1}$.
Since $\O(\f)$ and $K_{ij}(\f)$ are independent of $b$,
the integral,
\eqn\bidntz{\eqalign{
0 & = \int \cD\f  \cD b  \cD \psi  \cD \pb
\ \O(\f(0)) \  K_{ki}^{-1}(\f(0)) \ { {\d \exp I_{\rm tot}} \over {\d
b_i(0)} } \ ,
}}
vanishes because the integrand is a derivative with respect to
$b$.  By \eact, this gives
\eqn\bidnt{\eqalign{
0 & = \langle \  \O \  K_{ki}^{-1} \
\Big( \ \dot{\f}_i + K_{ij}{ {\p S} \over {\p \f_j} } + 2K_{ij}b_j
- \pb_j{ {\p K_{ji}} \over {\p \f_l} } \psi_l \ \Big) \ \rangle_{d+1} \ ,
\cr
 & = \langle \  \O \
\Big( \ K_{ki}^{-1} \  \dot{\f}_i + { {\p S} \over {\p \f_k} } + 2b_k
- K_{ki}^{-1} \ \pb_j{ {\p K_{ji}} \over {\p \f_l} } \psi_l \ \Big) \
\rangle_{d+1} \ ,
}}
where we have written $\O = \O(\f(0))$ etc, and it is understood that all
fields are
evaluated at $t = 0$. We will show in the next section that the reduced action
$I_{\rm tot}'$, obtained  after integrating out the ghosts, is invariant
under a time reversal
transformation under which $\f$ transforms according to $\f_i(t) \to
\f_i(-t)$, so
\eqn\oddt{\eqalign{
	\langle \  \O(\f(0)) \ K_{ki}^{-1}(\f(0)) \  \dot{\f}_i(0) \
\rangle_{d+1} = 0 .
}}
We write $b_j = s\pb_j$, and use $s$-invariance to obtain
\eqn\aidnt{\eqalign{ 0 & = \langle \  \O \  \Big(
 \ { {\p S} \over {\p \f_k} }
+ K_{ki}^{-1}{ {\p K_{ij}} \over {\p \f_l} } \psi_l \pb_j\ \Big)
	- 2s \O \  \pb_k \  \rangle_{d+1} ,
\cr
 & = \langle \  \O { {\p S} \over {\p \f_k} }
	- 2 \ { {\p \O} \over {\p \f_l} }  \psi_l \pb_k
	+ \O K_{ki}^{-1} { {\p K_{ij}} \over {\p \f_l} } \  \ \psi_l \pb_j
\  \rangle_{d+1} .
}}
All fields are evaluated at $t = 0$.

	In Appendix A we show that inside the expectation value
we may make the substitution
\eqn\ghsub{\eqalign{
 \psi_j(0) \pb_l(0) \to \demi \d_{jl}.
}}
This gives
\eqn\eidnt{\eqalign{
0 = \langle \  \O { {\p S} \over {\p \f_k} }
	-   \ { {\p \O} \over {\p \f_k} }
	+ \demi K_{ki}^{-1}\O{ {\p K_{ij}} \over {\p \f_j} }   \
\rangle_{d+1} ,
}}
and by \kernel\ we obtain the SD equations of
the $d$-dimensional theory
\eqn\fidnt{\eqalign{
0 = \langle \  \O { {\p S} \over {\p \f_k} }
	-  \ { {\p \O} \over {\p \f_k} }  \  \rangle_{d+1} ,
}}
which hold for every $\O = \O(\f(0))$.
We have thus proven that the two formulations give the same correlation
functions, at the very basic level of Dyson--Schwinger equations.

\newsec{Time reversal, canonical structure, and stability of the action}

	In dealing with renormalizable theories, one must ask	what is the
most general action that is compatible with given symmetries and field
dimensions?  In particular do the symmetries assure that, under
renormalization, a topological $(d+1)$-dimensional action remains
equivalent to a local $d$-dimensional Euclidean theory?   With reference
to eqs. \tact\ and \eact, we see that symmetry must preserve the form of
term $b_iK_{ij}({ {\p S} \over {\p \f_j} } + b_j)$.
Counter-terms cannot be tolerated that would change it, for example, to the
form $b_iK_{ij}(M_j + b_j)$, where
$M_j(\f)$ is not derivable from an action.  Fortunately such counter-terms
are forbidden by the symmetry under time reversal that we already used in
the previous section.  This is a symmetry of the $\f$--$b$ sector that is
obtained by setting  to 0 all sources with non-zero ghost number.  The
symmetry of this sector is controlled by the symmetry of the reduced action
that is obtained by integrating out the ghost fields.

	{\it Statement}:  The reduced action $I'_{\rm tot}(b, \f)$, obtained by
integrating out the ghost fields, is local,
and invariant under the time reversal transformation
	\eqn\trev{\eqalign{
\f_i(t) & \to \f_i^{\rm T}(t) = \f_i(-t)    \cr
b_i(t) & \to b_i^{\rm T}(t) = -  b_i(-t) - { {\p S} \over  {\p \f_i} }(-t).
}}
In terms of the variable
$b_i' \equiv b_i + \demi  { {\p S} \over {\p \f_i} } $
which transforms according to
\eqn\treva{\eqalign{
b'_i(t) \to {b'_i}^T(t) =  - b'_i(-t),
}}
the reduced action is given by
	\eqn\effactp{\eqalign{
 I_{\rm tot}'(\f, b')
= \int dt  \
\{ b'_i \dot{\f}_i +  b'_i K_{ij}b'_j
	- \quart \  { { \p S} \over {\p \f_i} } K_{ij} { { \p S} \over {\p
\f_j} }
+ \demi K_{ij}
{ { \p^2S} \over {\p \f_j \p \f_i} } \ \}.
}}
This action has a standard canonical structure, with momentum canonical to
$\f_j$ given by $p_j = i b'_j$.  It is obviously invariant under the above
time-reversal transformation.  The reduced action and the
time-reversal transformation and the canonical change of variable from $b$
to $b'$ are all local.

{\it Proof}: By \eact, the integral over the ghosts gives
\eqn\intgh{\eqalign{ \int d\psi d\pb \exp I_{\rm gh} = \det [ \p/\p t +
L(\f, b)],
}}
where $L$ is given in \defl.   Apart from an irrelevant multiplicative
constant, this may
be written
\eqn\jxp{
\eqalign{ \int d\psi d\pb \exp I_{\rm gh} & = \det [ 1 + G_0L_{\rm int}(\f,
b)]  \cr
& = \exp \Tr \ln[ 1 + G_0L_{\rm int}(\f, b)]  \cr
& = \exp \Tr [ L_{\rm int}G_0
 - \demi L_{\rm int} G_0 L_{\rm int} G_0 + ...] .
}}
Here $G_0$ is the integral operator
$G_0 =(\p/\p t + L_0)^{-1}$, with kernel $G_{0,ij}(t-u)$, given in
(A.7).  This kernal
is retarded, $G_{0,ij}(t-u) = 0$ for $t < u$, so all terms in the expansion
vanish except the
first, and we have
\eqn\jloc{\eqalign{
\int d\psi d\pb \exp I_{\rm gh}  &
= \exp [\int dt \ L_{{\rm int},ij}G_{0,ji}(t-u)|_{u=t} ] \cr
 & = \exp [\int dt \ L_{{\rm int},ii} \t(0)] \cr
	& = \exp [\int dt \ \demi L_{{\rm int},ii} ] \ ,
}}
where $\t(t)$ is the step function, and we have used the consistent
determination $\t(0) =
\demi$.  By \defl\ and \kernel, this gives
\eqn\jloc{\eqalign{
\int d\psi d\pb \exp I_{\rm gh} = \exp ( \int dt \ \demi K_{ij}
{ { \p^2S} \over {\p \f_j \p \f_i} } ) \ .
}}
Only the tadpole term survives, and the ghost integral contributes a
local term to the reduced action that depends only on $\f$.  From
 \eact, we conclude that the reduced action, obtained by
integrating out the ghosts, is given by
\eqn\effact{\eqalign{
I_{\rm tot}'(\f, b)  = \int dt \ \{ \  \demi \dot{S} + b_i [\dot{\f}_i
+ K_{ij}( { {\p S} \over {\p \f_j} } + b_j)] + \demi K_{ij}
{ { \p^2S} \over {\p \f_j \p \f_i} } \ \}  \ .
}}
With $\dot{S} =  { {\p S} \over {\p \f_i} } \dot{\f}_i$, eq. \effactp\
follows.

	For completeness we exhibit the integration over the canonically
conjugate
field $b'$, which is immediate from \effactp.   The integral is Gaussian and
converges because $b'$ is a purely imaginary field,
\eqn\effactb{\eqalign{
\exp I_{\rm tot}''(\f) \equiv \int db' \ \exp I_{\rm tot}'(\f, b')
= (\det K)^{-1/2} \exp \int dt  \
\{ & -\quart \dot{\f}_i (K^{-1})_{ij}\dot{\f}_j   \cr
	- \quart \  { { \p S} \over {\p \f_i} } K_{ij} { { \p S} \over {\p
\f_j} }
&+ \demi K_{ij}
{ { \p^2S} \over {\p \f_j \p \f_i} } \ \}.
}}

It should be noticed that when the kernel $K_{ij}$ is the identity, the time
reversal symmetry extends to the ghost part of the action, with the
transformation
$\psi_i(x,t)\to \bar \psi_i(x,-t)$ and $\bar \psi_i(x,t)\to \psi_i(x,-t)$.
For a generic kernel the ghost part of the action is not invariant under time
reversal.  However, as we shall show shortly,
lack of time-reversal invariance in the ghost sector does not prevent us
from using time-reversal invariance in the $\f$--$b$ sector
to prove stability of the action  under renormalization and
convergence of the correlation functions in a slice at fixed time toward those
of the ordinary formulation. The lack of the time reversal invariance in the
ghost sector seems to  be an artifact of the choice of a kernel, and
cannot  affect physical quantities.


Symmetry under the time reversal selects the $(d+1)$-dimensional
theories that are derivable from a $d$-dimensional action $S$.
Indeed, if the counter-terms generated a generic drift force that is not
derivable from an action,
$M_j(\f) \neq { {\p S} \over {\p \f_j} }$, then
upon integrating out the ghost fields and completing the square
as in \effactp, one gets the cross term
$(-\demi) \int dt \dot{\f}_i M_i(\f)$.  This term is not an exact time
derivative, and would violate time-reversal
invariance unless $M_j(\f) = { {\p S} \over {\p \f_j} }$, for some
Euclidean action $S$, in which case it is a mere boundary term
$(-\demi) \int dt \dot{S}$. More precisely, if we consider the class of
actions
\eqn\effactb{\eqalign{
s (\bar \Psi _i (\pa_t \phi_i +P_i(\phi) +K_{ij} b_j)),
}}
one gets the $t$-parity violating term
$(-\demi)\int dt \dot{\f}_i K_{ij}^{-1} P_j(\f)$
which disappears if and only
if the integrand is a boundary term, that is when $ P_i(\f)=K_{ij} { {\p S}
\over {\p \f_j} }$, which is the desired property. Part of our intuition is
actually that the integral
$\int dt \dot{\f}_i { {\p S} \over {\p \f_j} }$ gives back the action $S$
in $d$ dimensions, so we might interpret the quantization with an additional
time as a very refined version of Stokes theorem, generalized  to the case of
path integration.

 We now sketch the argument that establishes the stability of the action
in the context of renormalizable theories.  The full renormalized action must
be $s$-exact, and the first step is to construct the most general local
renormalized action that is $s$-exact and allowed by power counting.  Next
we use the important property of invariance under inversion of the 5th time
which is a property of the correlation functions in the $b$--$\f$ sector,
namely correlation functions with external $\f$ and $b$ legs only.
Implementation of this symmetry is facilitated by that fact that the
renormalization of correlation functions in the $\f$--$b$ sector
does not involve the ghost renormalization because  the ghost propagators
are retarded and cannot form closed loops, apart from the tadpole term
which is local.  (The tadpole term vanishes with dimensional regularization).
The $t$-reversal symmetry leaves no room for a drift force that is not a
gradient, so the ``gauge function'' must remain of the form $b_j+ { {\p S}
\over {\p \f_j} }$, where $S$ is a renormalized action.  Finally, BRST
symmetry is used to determine the ghost part of the renormalized action
from the $\f$--$b$ part.  We conclude that the combination of both
time-reversal symmetry and BRST invariance  implies  that the renormalized
$(d+1)$-dimensional theory remains equivalent to a local $d$-dimensional
theory on a time-slice.

	Note:  Because the time-reversal transformation
$b(t) \to - b(-t) - {{\p S}\over {\p \f}}$, is non-linear, one must
introduce a source term $\int dt N_i{{\p S}\over {\p \f_i}}$
in order to maintain this symmetry in the context of a renormalizable
theory.  (For BRST invariance one also introduces a source for
$s{{\p S}\over {\p \f_i}} = {{\p^2 S}\over {\p \f_i \p \f_j }}\psi_j$.)
The reduced partition function
$Z_1(j_\f, j_b, N) = \exp W_1(j_\f, j_b, N)$,
obtained by setting to 0 all sources with non-zero ghost number,
may be expressed in terms of the reduced action $I'_{\rm tot}(\f, b)$,
\eqn\redpart{\eqalign{
Z_1(J_\f, J_b, N)
= \int d\f db \  \exp[I'_{\rm tot}(\f, b)
+  \int dt  \ (j_{\f_i} \f_i + j_{b_i} b_i + N_i {{\p S}\over {\p \f_i}})].
}}
It is easy to see that invariance of $I'_{\rm tot}(\f, b)$
under the non-linear time reversal transformation
$b(t) \to - b(-t) - {{\p S}\over {\p \f}}(-t)$ [and $\f(t) \to \f(-t)$]
implies that the generating functional of connected
correlation functions satisfies the symmetry condition
\eqn\redsym{\eqalign{
W_1(j_\f, j_b, N) = W_1(j_\f^T, j_b^T, N^T + j_b^T),
}}
where $j_\f^T(t) = j_\f(-t)$, $j_b^T(t) = -j_b(-t)$, and
$N^T(t) = N(-t)$.
However it does not appear that this condition is easily expressed in terms
of the reduced effective action $\G_1(\f, b, N)$ obtained from
$W_1(j_\f, j_b, N)$ by Legendre transformation.
The solution is to
introduce a source $j_{b'_i}$ for the canonically conjugate
field $b'_i = b_i + \demi {{\p S}\over {\p \f_i}}$,
which obeys the elementary transformation law
$b'(t) \to - b'(-t)$.  The corresponding reduced partition function
is defined by
\eqn\redparta{\eqalign{
Z'(j_\f, j_{b'}, N) & = \exp W'(j_\f, j_{b'}, N)    \cr
& \equiv \int d\f db' \ \exp[ \ I'_{\rm tot}(\f, b')
+  \int dt  \ (j_{\f_i} \f_i + j_{b'_i} b'_i) + N_i {{\p S}\over {\p \f_i}}
\ ] ,
}}
where $I'_{\rm tot}(\f, b') $ is the local action \effactp.  The two generating
functionals are related by
$W'(j_\f, j_{b'}, N) = W_1(j_\f, j_{b'}, N + \demi j_{b'})$, so they provide
the same information.  Invariance of $I'_{\rm tot}(\f, b')$  under the
elementary time-reversal transformations of $\f$ and $b'$ implies that
$W'(j_\f, j_{b'}, N)$ satisfies the elementary and standard time-reflection
condition $W'(j_\f, j_{b'}, N) = W'(j_\f^T, j_{b'}^T, N^T)$,
where $j_\f^T(t) = j_\f(-t)$, $j_{b'}^T(t) = -j_{b'}(-t)$, and
$N^T(t) = N(-t)$.  Consequently the reduced effective action
$\G'(\f, b', N)$, obtained by Legendre transformation from
$W'(j_\f, j_{b'}, N)$, also satisfies the elementary and standard
time-reflection condition
\eqn\redsym{\eqalign{
\G'(\f, b', N) = \G'(\f^T, b'^T, N^T),
}}
where $\f^T(t) = \f(-t)$, $b'^T(t) = - b'(-t)$ and $N^T(t) = N(-t)$.  In each
order of perturbation theory, the divergent terms  in the $\f$--$b$ sector are
local contributions to $\G'(\f, b', N)$ which must satisfy this standard
symmetry condition.

\newsec{S-matrix}

	We have shown above that the correlation functions of the
{d}-dimensional theory are
obtained from the {(d+1)}-dimensional theory by
\eqn\frfv{\eqalign{
G_{d}^{(n)}(x_1,... \ x_n) = G_{d+1}^{(n)}(t_1,x_1,... \ t_n, x_n)|_{t_1 = ...
t_n = 0} \ ,
}}
where $G_{d}^{(n)}$ and $G_{d+1}^{(n)}$ are correlators of $n$
$\f$-fields.  We write this in 5-dimensional momentum space as
\eqn\frfvm{\eqalign{
G_{d}^{(n)}(p_1,... p_n) = (2\pi)^{-n+1} \int \prod_{i=1}^n dE_i
\ \d( \sum_{i=1}^n E_i) \ G_{d+1}^{(n)}(E_1,p_1,... E_n, p_n) \ .
}}
Here the $\d$-function $\d(\sum_i E_i)$ expresses invariance under
translation in the
{(d+1)}-th time.  We have removed a momentum conserving $\d$-function
$\d^{d}(\sum_{i=1}^n p_i)$ from both sides, so the $p_i$ are understood to be
constrained by $\sum_{i=1}^n p_i=0$.  The {(d+1)}-dimensional action
allows one to derive Feynman rules for the {(d+1)}-dimensional
correlation functions in the Euclidean region, where the $p_\m$ are real for
$\m = 1,...{d}$.  In the following we shall deal with the connected
components of the correlators and of the S-matrix which alone may be
continued analytically.  However we shall not trouble to write out the
expansion in terms of connected components explicitly.

	To obtain the physical S-matrix, the $(d+1)$-dimensional
connected correlators
$G_{d+1}^{(n)}(E_i,p_i)$
must be continued from the Euclidean region in the
$p_{i,\m}$ , where $\m = 1,...d$, to the on-shell Minkowskian region,
$p_i^2 = - m_i^2$, for real positive mass, $m_i \geq 0$, by continuation to
imaginary values of the $d$-th component $p_{i,d}$. We shall see below
that the region of analyticity of the $(d+1)$-dimensional correlators in the
$p_{i,\m,}$ is at least as great as it is for the $d$-dimensional
correlators, so this continuation is always possible.  According to the
LSZ method in $d$ dimensions, the S-matrix is obtained from the
correlation function in momentum space by amputating each leg
namely by multiplying by
$X_i \equiv (p_i^2 + m_i^2)$ and going on mass shell.  Thus
the S-matrix element with $n$ external legs is given by
\eqn\lsz{\eqalign{
S^{(n)} & = \lim_{X_i \to 0}  \prod_{i=1}^n X_i  \ G_{d}^{(n)}(p_i) \cr
S^{(n)} & = (2\pi)^{-n+1}\lim_{X_i \to 0}  \prod_{i=1}^n X_i
\int \prod_{i=1}^n d E_i  \ \d(\sum_{i=1}^n E_i) \ G_{d+1}^{(n)}(E_i, p_i)\ .
}}

	We wish to express $S^{(n)}$ directly in  terms of the truncated
{(d+1)}-dimensional
correlation functions $\G_{d+1}^{(n)}$ which are related to the
$G_{d+1}^{(n)}$ by
\eqn\trunk{\eqalign{
(G_{d+1}^{(n)})_{\f_1,... \ \f_n} = D_{\f_1 \a_1} ...  \ D_{\f_n \a_n}
\ (\G_{d+1}^{(n)})_{\a_1,... \ \a_n} \ ,
}}
where $\a_i \equiv (\f_i,b_i)$ is a 2-valued index.  We write this in matrix
notation as
\eqn\trunkm{\eqalign{
G_{d+1}^{(n)} = \prod_i D_i \  \G_{d+1}^{(n)},
}}
where it is understood that the row label of $D_i$ is $\f$.
Here $D_{\a \b} \equiv D_{\a \b}(E, p)$
is the $2 \times 2$ propagator matrix,
\eqn\prop{\pmatrix{
D_{ \f \f }(E, p) &
D_{ \f b }(E, p)     \cr
D_{ b \f }(E, p) &
0
}}
and $D_{ b b}(E, p) = 0$ by $s$-invariance, as shown in (4.10) below.
This gives
\eqn\tlsz{\eqalign{
S^{(n)}  = (2\pi)^{-n+1}\lim_{X_i \to 0}   \int \prod_{i=1}^n d E_i
\ \d(\sum_{i=1}^n E_i) \ \prod_{i=1}^n (X_i D_i) \ \G_{d+1}^{(n)}(E_i, p_i)\ ,
}}
Because each leg is
multiplied by $X_i \to 0$, only a singular part will survive.

We consider the case of scalar particles with kernel $K = 1$ in which case the
free propagators in ``energy'' and momentum space are given by
\eqn\freeprop{\eqalign{
(D_0)_{ \f \f } & = 2[E^2 + (p^2 + m^2)^2]^{-1} \ \ \ \ \ \ \
(D_0)_{ \f b } = - (p^2 + m^2 + iE)^{-1}    \cr
(D_0)_{ b \f } & = - (p^2 + m^2 - iE)^{-1}  \ \ \ \ \ \ \ \ \
(D_0)_{ b b} = 0.
}}
The external ``energies" are real and non-zero, so the domain of
analyticity in the external momentum components
$p_\m$ is greater than for the corresponding $d$-dimensional
Euclidean Feynman graphs, and thus the continuation from the Euclidean to
the on-shell Minkowskian region is possible.  We will also show below that
the relevant graphs are analytic in every upper-half $E_i$-plane, where
$E_i$ is the ``energy" of the i-th external line.

	To proceed further we use simple properties of truncated diagrams
which are a
direct consequence of the fact that the bulk action is $s$-exact.  From the
elementary identities
\eqn\sidtty{\eqalign{
0 & =  \langle \  s\big(\pb(t,x) b(0,0) \big) \ \rangle
= \langle \ b(t,x) b(0,0) \ \rangle \cr
0 & =  \langle \  s \big(\f(t,x) \pb(0,0) \big) \ \rangle
= \langle \ \psi(t,x) \pb(0,0) \ \rangle
+ \langle \ \f(t,x) b(0,0) \ \rangle \
}}
we obtain
\eqn\propi{\eqalign{
D_{bb}(t, x) = 0     \ \ \ \ \ \ \ D_{ \f b }(t, x) = - D_{ \psi \pb }(t,
x) \ .
}}
Moreover  the ghost propagator $D_{ \psi \pb }(t, x)$ is retarded,
as shown in Appendix A, and thus so is $D_{ \f b }$,
\eqn\retard{\eqalign{
D_{ \f b }(t, x) = - D_{ \psi \pb }(t, x) = 0  \ {\rm for}  \ t < 0.
}}
There are no closed
ghost loops, apart from a tadpole term, because the ghost propagator is
retarded.

	We now use these properties to show that every non-zero truncated
diagram
contains at least one external $b$-line.  Note from the action \eact\ with
$K =1$ that each vertex contains precisely one $b$-line.  Now consider a fixed
vertex of the diagram.  Call it $V_1$.  Either its $b$-line is external,
and if so the
assertion is true, or if not, this $b$-line connects to another vertex, call it
$V_2$, by a $(D_0)_{\f b}$ propagator because the $(D_0)_{b b}$ propagator
vanishes.  We now repeat the argument for $V_2$.  The vertex $V_2$ contains
one $b$-line which is either external or connected to a third vertex  $V_3$
by a
$(D_0)_{\f b}$ propagator.  (See Fig. 1.)  Because the $(D_0)_{\f b}$
propagators are all retarded, this sequence follows the direction of increasing
time, and each vertex in the sequence is different.  We conclude that every
finite
diagram has at least one external $b$-line, as asserted.   One may also
prove this
algebraically using the Ward identities for $s$-invariance.   They
imply that the one-particle irreducible correlation functions have this
property,
from which it follows that the truncated correlation functions do.

	The above argument also shows that every
truncated diagram $\G_{d+1}$ vanishes unless the largest time is associated
to an
external $b$-line.  In particular, if there is only a single external
$b$-line, then it must be associated to the largest time.  In fact we will see
shortly that only truncated graphs with a single $b$-line contribute to the
S-matrix.  Consider one such graph, and let the
$\f$-fields be at $\f(x_i, t_i)$, for  $i = 1,..., n-1$, and let the
$b$-field be taken at
the origin $b(0,0)$.  This graph vanishes if any one of the $t_i$ is positive.
Therefore, in momentum space, the graph in question is analytic in every
upper-half $E_i$-plane.

  Let $\G_{d+1}^{(n-p,p)}$ be the part of $\G_{d+1}^{(n)}$ with $p$
external $b$-lines and $(n-p)$ external $\f$-lines, and call
$S^{(n-p,p)}$ the
corresponding contribution to $S^{(n)}$, according to \tlsz.  We have just
shown that
$\G_{d+1}^{(n,0)} = 0$, so $S^{(n)} = \sum_{p = 1}^n S^{(n-p,p)}$.  We will
see shortly that
$S^{(n-p,p)} = 0$ for $p \geq 2$, so
\eqn\ssone{\eqalign{
S^{(n)} = S^{(n-1,1)} \ .
}}
We next evaluate
$S^{(n-1,1)}$ which comes from $\G_{d+1}^{(n-1,1)}$.  Any single one of the $n$
external lines of $\G_{d+1}^{(n-1,1)}$ may be the $b$-line, and we have
$\G_{d+1}^{(n-1,1)} =\sum_{j = 1}^n\G_{d+1}^{(n-1,1,j)}$, where
$\G_{d+1}^{(n-1,1,j)}$ is
the truncated correlation function where the $j$-th external line is a
$b$-line, and
all other lines are $\f$-lines.  Correspondingly we have the decomposition
\eqn\exps{\eqalign{
S^{(n)} = S^{(n-1,1)} = \sum_{i = 1}^n S^{(n-1,1,j)} \ .
}}

	 We now evaluate
 \eqn\sone{\eqalign{
S^{(n-1,1,j)} =   (2\pi)^{-n+1} &
\int d E_j
\lim_{X_j \to 0} X_j D_{\f b}(E_j, p_j)  \cr
   & \int \prod_{i \neq j} d E_i  \lim_{X_i \to 0}X_i \  D_{\f \f}(E_i, p_i) \
	\d(\sum_{i =1}^n E_i)\G_{d+1}^{(n-1,1,j)}  \ ,
}}
where $X_i = p_i^2 + m_i^2$.  Integration with respect to $E_j$
eliminates the $\d$-function,
$\int d E_j \d(\sum_{i=1}^n E_i) = 1$, and we have the constraint that the
``energy" is conserved.  The remaining external lines, for $i \neq j$ are all
$\f$-lines. We must calculate $	\lim_{X_i \to 0} X_i  D_{\f \f}(E_i, X_i)$,
where
$X_i = p_i^2 + m^2$.  Let us provisionally make the simplifying assumption that
$D_{\f \f}(E_i, X_i)$ and  $D_{\f b}(E_j, X_j)$ are
the free propagators \freeprop.  In this case we have the
simple and useful limit
 \eqn\deltaf{\eqalign{
\lim_{X_i \to 0} X_i  D_{\f \f}(E_i, X_i)
= \lim_{X_i \to 0} { {2X_i} \over {E_i^2 + X_i^2} } = 2\pi \d(E_i) .
}}
We use this identity for all $i \neq j$, so when we integrate on $E_i$ we
have
$	\int dE_i \d(E_i) = 1$, and everywhere else we set $E_i = 0$
and $X_i = 0$ for
$i \neq j$.  By conservation of ``energy" we also have $E_j = 0$.  We have
provisionally assumed that $D_{\f b}(E_j, X_j)$ is the free propagator
\freeprop, so
\eqn\phib{\eqalign{
\lim_{X_j \to 0} X_j D_{\f b}(E_j, p_j)|_{E_j = 0}
= \lim_{X_j \to 0} \Big({ {X_j} \over {X_j + iE_j} }|_{E_j = 0}\Big) = 1 \ ,
}}
and everywhere else we set $X_j = 0$.  This gives
 \eqn\redf{\eqalign{
S^{(n-1,1,j)} =  \ 	\G_{d+1}^{(n-1,1,j)}|_{E_i = 0, X_i = 0}
}}
for all $i = 1,... \ n$.  Since the external $b$-leg may be any one of the
legs we
obtain the remarkably simple
formula
 \eqn\redfs{\eqalign{
S^{(n)} = S^{(n-1,1)}
=  \ \sum_{j = 1}^n	\G_{d+1}^{(n-1,1,j)}|_{E_i = 0, X_i = 0}  \ .
}}

	We must correct the above argument by use of the exact
$D_{\f \f}(E, p)$ and $D_{\f b}(E, p)$
propagators instead of the free ones.  The exact propagator is a $2 \times
2$ matrix
which is the inverse of the matrix
\eqn\proper{\pmatrix{
   0 &    \G_{\f b}(E, p)    \cr
	\G_{b \f}(E, p)  &  \G_{b b}(E, p)
}}
where $\G_{\f b}(E, p) =   \G_{b \f}(- E, - p) = \G_{b \f}(- E, p),
= \G_{b \f}^*( E, p)$
namely
\eqn\dfb{\eqalign{
D_{\f b}(E, p) = { {1} \over {\G_{b \f}(E, p)} }
}}
\eqn\dff{\eqalign{
	D_{\f \f}(E, p) = { {- \G_{b b}(E, p)}  \over { | \G_{b \f}(E, p)
|^2 } }.
}}
These quantities have the perturbative expansions
\eqn\devp{\eqalign{ -	\G_{b b} =  1  +  O(g^2) }}
\eqn\devpp{\eqalign{	- \G_{b \f}  =  i E + p^2 + m_0^2 + O(g^2),}}
where $m_0$ is the bare mass, as one reads off from
$\G_0 = - I_0$, where $I_0$ is the zeroth order part of the action $I$.
To zeroth order $\G_{b \f}(E, p)$ vanishes at
$i E + p^2 + m_0^2 = 0$.  We suppose
that the exact quantity $\G_{b \f}(E, p)$ vanishes at
\eqn\zofe{\eqalign{
iE + f(p^2) = 0,
}}
where $f(p^2)$ is an arbitrary real function.  Then we have
\eqn\pole{\eqalign{
- \G_{b \f}(E, p) = [ i E + f(p^2) ] R(E, p^2),
}}
where $R(E, p^2)$ is regular at $iE + f(p^2) = 0$.   We suppose also that
$f(p^2)$ vanishes at $	p^2 = - m^2$, where, naturally, $m$ is the renormalized
mass.    We define
$	X \equiv p^2 + m^2$, and we have
\eqn\ppole{\eqalign{
f(p^2) = X r(X),
}}
where $r(X)$ is a function that is regular at $X = 0$.  This gives
\eqn\gpole{\eqalign{
\G_{b \f}(E, p) = [ i E + X r(X) ] R(E, X),
}}
\eqn\dpole{\eqalign{
D_{\f b}(E, p) & = { {1}  \over {[ i E + X r(X) ] R(E, X)} }  \cr
D_{\f \f}(E, p) & = { {- \G_{b b}(E, X)} \over
{| R(E, X) |^2	   [ E^2  +  X^2 \ r^2(X) ]} } .
}}
Instead of \deltaf\ we now obtain
 \eqn\deltag{\eqalign{
\lim_{X_i \to 0} X_i  D_{\f \f}(E_i, X_i)
= 2\pi Z_{\f \f} \d(E_i) .
}}
where $Z_{\f \f}$ is the renormalization constant
\eqn\rcon{\eqalign{
Z_{\f \f} =  { {- \G_{b b}(0, 0)} \over {| R(0, 0) |^2  r(0)} } \ ,
}}
and instead of \phib\ we obtain
\eqn\phic{\eqalign{
\lim_{X_j \to 0} X_j D_{\f b}(E_j, p_j)|_{E_j = 0}
= Z_{\f b} \ ,
}}
where
\eqn\reconb{\eqalign{
Z_{\f b} = { {1} \over {r(0)  R(0, 0)} } \ .
}}
Thus, apart from renormalization constants, we obtain \redfs.

	There remains only to show that the contribution to the S-matrix from
$\G_{d+1}^{(n-p,p)}$ vanishes for $p \geq 2$, namely if there is more than one
external
$b$-line on
$\G_{d+1}^{(n)}$.  But this is immediate because we have
\eqn\reg{\eqalign{
\lim_{X \to 0}  X D_{\f b}(E, p)
 = \lim_{X \to 0}  { {X}  \over {[ i E + X r(X) ] R(E, X)} }
           =  0,
}}
since, in contradistinction to \phib, E is not contrained to be 0.

	We conclude that the S-matrix is given by \redfs\ corrected by the
renormalization constants, namely
\eqn\rega{\eqalign{
S^{(n)}(p_1,... \ p_n) = Z_{\f b} \ Z_{\f \f}^{n-1}  \sum_{j=1}^n
 \G_{d+1}^{(n-1, 1, j)}(E_1, p_1,... \ E_n, p_n)|_{E_i = 0, X_i = 0} \ .
}}
Thus in the bulk formalism, the S-matrix is obtained from $\G_{d+1}^{(n-1,
1,j)}$, the
truncated correlation function with exactly one external $b$-line (in all
possible
positions $j$) by going ``on-shell", namely by setting $E_i = 0$ and
$X_i=0$ for all $i$.

\newsec{Landau-Cutkosky rules and perturbative unitarity}

	The Landau-Cutkosky rules follow from properties of integrals of
the form $\int dx f(x, z)$, where $x$ and $z$ are sets of complex variables,
corresponding respectively to internal and external {(d+1)}-momenta.
These rules hold wherever these variables may be continued.  Thus to derive
the Landau-Cutkosky rules, we may use Feynman rules for the Euclidean
correlators, with the understanding that a continuation in external momenta
is made to the Minkowski region, as discussed in the last section.

	Consider an on-shell connected and truncated graph which
contributes to the
S-matrix.  As we have just shown, it has
an arbitrary number of external $\f$-lines, and precisely one external
$b$-line.
We cut it into two subgraphs across $n$ intermediate lines, with incoming and
outgoing particles on opposite sides of the cut.   Let
$p_i$  and $E_i$, for  $i = 1,...n$, be the ``momentum" and ``energy" of the
$i$-th
intermediate line.  The total $d$-momentum
$P = \sum_{i=1}^n p_i$ entering through the incoming
lines is fixed, as is the total ``energy''
which vanishes, since it is on-shell,
$0 = \sum_{i=1}^n E_i$.  Thus the integration associated with the cut lines
is
\eqn\intgrl{\eqalign{
\prod_{i=1}^{n} d^{d}p_i dE_i \
\d^{d}(\sum_{i=1}^n p_i - P) \d( \sum_{i=1}^n E_i)
= \prod_{i=1}^{n-1} d^{d}p_idE_i \ .
}}

	The one external $b$-line of the S-matrix falls on one side of the
cut, and we consider a particular
vertex on the other side of the cut.  As was noted above, we may continuously
follow $(D_0)_{\f, b}$ propagators from any vertex of the graph to the one
external $b$-line.  See Fig. 2. Therefore at least one of the intermediate cut
lines is a
$(D_0)_{\f, b}$ propagator.  Consider first the case when $n-1$
intermediate cut lines are $(D_0)_{\f, \f}$ propagators
\eqn\phiphi{\eqalign{
2[E_i^2 + (p_i^2 + m_i^2)^2]^{-1}
= 2(p_i^2 + m_i^2 + iE_i)^{-1}(p_i^2 + m_i^2 - iE_i)^{-1}
}}
for $i = 1,... (n-1)$, while the $n$-th line is a $(D_0)_{\f, b}$
propagator
\eqn\phibe{\eqalign{
- (p_n^2 + m_n^2 + iE_n)^{-1}
 = - [(P - \sum_{i=1}^{n-1}p_i)^2+m_n^2 + i(-  \sum_{i=1}^{n-1}E_i )]^{-1} \ .
}}
Each subgraph has exactly one external $b$-line, so each contributes to an
S-matrix element by the reduction formula of the preceding section.  We wish to
find the location of the leading singularity associated with a pinching of the
integration \intgrl\ aassociated with these lines, and to calculate the
discontinuity
across it.

	For this purpose we combine all the denominators that correspond to cut
lines by means of Feynman's auxiliary parameters, using the factorized
form \phiphi\ for
the $(D_0)_{\f, \f}$ propagators.  This gives the overall denominator $D^n$,
where
\eqn\denom{\eqalign{
D(p_i, E_i, \a_i, \b_i, \a_n) \equiv \sum_{i = 1}^{n-1} [\a_i (p_i^2 +
m_i^2 + iE_i)
+ \b_i(p_i^2 + m_i^2- iE_i)]   \cr
+ \g [(P - \sum_{i=1}^{n-1}p_i)^2 + m_n^2 -  \sum_{i=1}^{n-1}iE_i] \ .
}}
According to the Landau rules, a pinching singularity occurs
where $D$ and all its first derivatives vanish,
\eqn\landaua{\eqalign{
{ {\p D} \over {\p \a_i} } = p_i^2 + m_i^2 + iE_i = 0
}}
\eqn\landaub{\eqalign{
{ {\p D} \over {\p \b_i} } = p_i^2 + m_i^2 - iE_i = 0
}}
\eqn\landauc{\eqalign{
{ {\p D} \over {\p \g} } = (P - \sum_{i=1}^{n-1}p_i)^2 + m_n^2
- \sum_{i=1}^{n-1}iE_i = 0 }}
\eqn\landaud{\eqalign{
{ {\p D} \over {\p p_i} } = 2(\a_i + \b_i)p_i
+ 2 \g( \sum_{i=1}^{n-1}p_i - P) = 0
}}
\eqn\landaue{\eqalign{
{ {\p D} \over {\p E_i} } = i(\a_i - \b_i - \g) = 0 \ ,
}} for $i = 1,...(n-1)$.  From \landaua\ and \landaub\ we obtain $E_i =
0$, and
$p_i^2 + m_i^2 = 0$ for $i = 1,...(n-1)$.  Together with \landauc, this
gives
$ (P - \sum_{i=1}^{n-1}p_i)^2 + m_n^2 = 0$.  So all $n$ lines are on
the mass-shell, and all $n$ energies vanish.  This implies that $D = 0$.
>From \landaue\ we obtain
$\b_i = \a_i - \g$.  We substitute this into \landaud\ and obtain
$(2\a_i - \g)p_i + \g( \sum_{i=1}^{n-1}p_i - P) = 0$.  We
change variable from $\a_i$ to $\a_i' \equiv 2\a_i - \g$, and we write
$\a_n' \equiv \g$, so the last equation reads
$\a_i'p_i + \a_n'( \sum_{i=1}^{n-1}p_i - P) = 0$.

	We conclude that the conditions for a singularity are the conditions
$E_i = 0$, plus the standard Landau equations for the momenta $p_i$
and $p_n$, which imply that all momenta are on the mass shell.  This is the
on-shell condition of the $(d+1)$ dimensional theory which we found for the
S-matrix.  The singularities occur at Minkowksian values of the $p_i$, and
the singularity in $P$ occurs at the physical threshold $P^2 = (\sum_{i=1}^n
m_i)^2$.

 To evaluate the discontinuity across the
cut that begins at $P^2 = (\sum_{i=1}^n m_i)^2$, we integrate over the
$E_i$, for $i = 1,... \ (n-1)$, by closing the contour in the upper
half-plane.  In this way
we pick up the $2\pi$ times the residue of the first pole in
\phiphi, at $E_i = i(p_i^2 + m_i^2)$.  As a result, each propagator
$(D_0)_{\f \f}(E_i, p_i)$ is
replaced by
$(p_i^2 +m_i^2)^{-1}$ for $i = 1,... (n-1)$, and the  propagator
$(D_0)_{\f b}(E_n, p_n)$ is replaced by
$[(P - \sum_{i=1}^{n-1}p_i)^2+m_n^2 +  \sum_{i=1}^{n-1}(p_i^2 + m_i^2
)]^{-1}$.
We have found from our Landau rules that the leading singularity occurs
when all these
propagators vanish, namely at
$p_i^2 + m_i^2 = 0$ for $i = 1,..., n$.  The problem of calculating the
discontinuity
has now been reduced to the familiar $d$-dimensional problem, and we
know from the $d$-dimensional Cutkosky rules that the leading discontinuity
is obtained
by replacing each propagator
$(p_i^2+m_i^2)^{-1}$ by $2\pi \d(p_i^2 + m_i^2)$.  The
net result is that the leading discontinuity is obtained by replacing each
propagator $(D_0)_{\f \f}(E_i, p_i)$ by
${d} \pi^2 \d(E_i) \d(p_i^2 +m_i^2)$ for $i = 1,...(n-1)$, and the propagator
$(D_0)_{\f b}(E_n, p_n)$ by
$2 \pi \d(p_n^2 +m_n^2)$.  This gives the discontinuity required for the
S-matrix
to be unitarity.

	Suppose now that more than one intermediate cut line is a
$(D_0)_{\f b}$ propagator.  Then the Landau rules that correspond to \landaua\
through \landaue\ are such that not all intermediate momenta are on mass
shell.
The corresponding pinching singularities cannot contribute to the unitarity
equation.

\newsec{Ward Identities for a global symmetry}

	We have shown that the topological $(d+1)$-dimensional
theory reproduces the standard $d$-dimensional theory in a time slice.
It follows that if the $d$-dimension action $S$ has a global invariance,
the Ward identities of the $d$-dimensional theory hold on a time-slice
of the $(d+1)$-dimensional theory.  However it is
possible to prove them directly in $d+1$ dimensions.
In the derivation which we present here, the Noether current in $d+1$
dimensions does not appear at all.  However in Appendix B  we give
an alternative derivation of the Ward identities that relies
on the $(d+1)$-dimensional Ward identity, with an
integration in $t$ on a very thin slice.  As in the derivation of the SD
equations, BRST-invariance plays an essential role.  At the end of this
section we explain how the anomalies of the $d$-dimensional theory
arise in the bulk formulation.

	Suppose that the infinitesimal transformation $\d \f = \e \l \f$  is a
symmetry of the $d$-dimensional Euclidean action $S$.  Here $\e$ is an
infinitesimal constant, and $\l$ is a numerical matrix that satisfies $\Tr
\l = 0$.  In this
case Noether's theorem holds, namely
${ {\d S} \over { \d \f(x)} } \l \f(x) = - \p_\m j_\m(x)$,
where $j_\m$, for $\m = 1,... d$, is the Noether current.  This
property of the action
leads to the Ward identity that holds in Euclidean quantum field theory
\eqn\dward{\eqalign{
0 = \langle
\Big (  {{\d  \O  }\over {\d  \phi(x)}} \l \phi(x)
+ \O  \p_\m j_\m(x)  \Big)\rangle_d \ .
}}
To establish this Ward identity in the bulk quantization, we start
from the identity
\eqn\nidnt{\eqalign{
0 & = \int \cD\f  \cD b  \cD \psi  \cD \pb
\ \O(\f(0)) \ (\l \f(0))_k K_{ki}^{-1}(\f(0))
\ { {\d \exp I_{\rm tot}} \over {\d b_i(0)} } \ ,
}}
where again the discrete index $i$ represents $x_\m$ and all internal
indices.  We repeat the steps of the derivation of the SD equations and
obtain
\eqn\widnt{\eqalign{
0 = \langle \  \O { {\p S} \over {\p \f_k} }(\l \f)_k
	-  \ { {\p \O} \over {\p \f_k} }(\l \f)_k  \  \rangle_{d+1} ,
}}
where all fields are evaluated at $t = 0$.  In continuum notation this reads
\eqn\xidnt{\eqalign{
0 = \langle \  \Big(\O { {\d S} \over {\d \f(x)} }\l \f(x)
	-  \ { {\d \O} \over {\d \f(x)} }\l \f(x)\Big)|_{\f(x) = \f(x,0)} \
\rangle_{d+1} .
}}
On using Noether's theorem ${ {\d S} \over { \d \f(x)} } \l \f(x) = - \p_\m
j_\m(x)$ for
the first term we recover the desired Ward identity.

	It may happen that the correlator
$\langle \p_\m j_\m(x,t) \ \O[\f(x,0)] \rangle_{d+1}$
 is discontinuous at $t = 0$, so that its value at $t = 0$ is
ambiguous.  In fact, as was shown in
\bgz, such a discontinuity does occur for $\O$ of the form
 $\O = j_a(y,0) j_b(z,0)$ when the 3 currents correspond
to the familiar triangle graph that exhibits the chiral anomaly. In this
case the above
Ward identity involving the $d$-dimensional Noether current $j_\m$ is
anomalous.
This happens even though the Ward identity for $\O =\O[\f]$
\eqn\emmaf{\eqalign{
0 & = \langle \ {{\delta^{d+1} \O }\over {\delta \phi(x,t)}} \l \f(x,
t) - \O \p _M K_M \ \rangle_{d+1} \ , }} is not anomalous, where $K_M$
for $M = 1,...(d+1)$, is the $(d+1)$-dimensional Noether current of
the bulk action $I_{\rm tot}$.  Here it is assumed that $I_{\rm tot}$
is invariant under the same symmetry as the Euclidean action $S$
provided that $\f$, $\psi$, $\pb$, and $b$ are transformed
appropriately.  The currents $j_\m$ and $K_M$ are different because
$S$ and $I_{\rm tot}$ are different.  The origin of the anomaly of
$j_\m$ in bulk quantization comes from a discontinuity of the
correlators at equal time.  It is perfectly admissible that the
$(d+1)$-dimensional Noether current $K_M$ is conserved, while the
$d$-dimensional Noether current $j_\m$ is not.

\newsec{Conclusion}

	We have shown the equivalence of the topological $(d+1)$-dimensional
formulation of quantum field theory to the standard $d$-dimensional
one for a theory of non-gauge type.   Our method does not rely on the
relaxation of a stochastic process and applies equally well to bose and
to fermi-dirac fields fields.  It thus provides a more general
framework than stochastic quantization.  It is also a more powerful
one, for we are able to show the stability of the topological
$(d+1)$-dimensional action that is equivalent to a local $d$-dimensional
theory.  We also expressed the physical $S$-matrix directly in
terms of the truncated correlation functions of $\f$ and its conjugate field
$b$ of the $(d+1)$-dimensional theory.  Since this approach goes
beyond the conceptual framework of stochastic quantization and does
not rely on the relaxation hypothesis, we call it ``bulk" quantization.
In a forthcoming publication we shall show that this approach allows
one to overcome the Gribov problem because, strictly speaking, one
does not really fix a gauge at all.  Instead the infinities associated with the
gauge modes are eliminated by making a gauge transformation in the
functional integral.

  \vskip .5cm
{\centerline{\bf Acknowledgments}}

The research of Daniel Zwanziger was partially supported by the National
Science Foundation under grant PHY-9900769.  The research of Laurent
Baulieu was supported in part by DOE grant DE-FG02-96ER40959.



\appendix A{Ghost propagator}

By \eact, the dependence of the action $I$ on the ghosts $\pb$ and $\psi$
is of the bilinear
form
\eqn\ghost{\eqalign{
I_{\rm gh} = - \int dt  \ \pb_i [ \d_{ij} { {\p} \over {\p t} } +
L_{ij}(\f, b) ] \psi_j \ ,
}}
where $L$ is given in \defl.  To calculate the ghost propagator with this
action we use the
identity
\eqn\ghid{\eqalign{
0 & = \int d\psi d\pb \ { {\d} \over {\d \pb_i(t)} } [\pb_k(u) \exp I_{\rm
gh}] \cr
	& =  \int d\psi d\pb \
[ \d_{ik} \d(t-u) + \pb_k(u)(\d_{ij}\p / \p t +L_{ij})\psi_j(t) ]
\exp I_{\rm gh} \ .
}}
Thus, with due attention paid to the anti-commutation of the ghost fields,
the ghost
propagator in fixed $\f$ and $b$ fields
\eqn\ghprop{\eqalign{
G(t,u;\f, b)_{jk} \equiv
N\int d\psi d\pb \  \psi_j(t) \pb_k(u) \exp I_{\rm gh}
}}
satisfies
\eqn\grfn{\eqalign{
[\d_{ij}\p / \p t+ L_{ij}(\f(t),  b(t))] G_{jk}(t,u;\f, b) = \d_{ik} \d(t-u).
}}

	For the cases of interest $\p / \p t + L$ is a parabolic operator.
Indeed $L$  is of the
form
$L(\f,b) = L_0 + L_{\rm int}(\f, b)$, where $L_0$ is independent of the
fields $\f$ and
$b$, and is a positive second order differential operator in $\p_\m$, for
$\m = 1,...d$.  For
example,
$L_0 = - \p^2 + m^2$, for the model \dact\ with $K_{ij} = \d_{ij}$, where
$\p^2 = \sum_{\m=1}^d \p_\m^2$.  (Recall that the discrete index $i$ or $j$
stands for
the continuum variables $x_\m$.)  We assume that $L_{\rm int}(\f, b)$
vanishes with
$\f$ and $b$ and contains no derivative $\p_\m$ higher than the first.
Because the
ghost propagator $G$ is the Green function of a parabolic operator, it is
retarded,
\eqn\ret{\eqalign{ G(t,u;\f, b)_{jl} = 0   \ \ \ {\rm for} \  t < u,
}}
(as follows from the integral equation below.)  This is a fundamental
property of the topological theory for it has as a consequence that all
closed ghost loops
vanish (except for the tadpole term which is given in \jloc).

	The free Green function $G_{0,ij}(t-u)$ is defined by
\eqn\grfno{\eqalign{
[\d_{ij}\p/\p t + L_{0,ij}] G_{0,jk}(t-u) = \d_{ik} \d(t-u),
}}
and is expressed in matrix or operator form by
\eqn\grfnp{\eqalign{
G_{0,ij}(t-u) = \t(t-u) \exp[ - L_0(t - u)].
}}
It is related to the exact Green function \grfn\ by
\eqn\rgrfn{\eqalign{
	G_{il}(s, u; \f, b) = G_{0,il}(s-u)
   - \int_u^s dt \ G_{0,ij}(s-t)L_{{\rm int},jk}(t)G_{kl}(t, u, \f, b).
}}
These expressions vanish unless $s \geq t \geq u$.  The second term
vanishes for
$s \to u$.  It follows from the last two equations that for $t \approx u$
\eqn\eqt{\eqalign{
G(t,u;\f, b)_{jl} = \d_{jl} \t(t - u) + o(t-u),
}}
where $\t(t-u)$ is the step function, and $o(t-u)$ vanishes with $t-u$.
Moreover a
consistent determination of $\t(0)$ is $\t(0) = 1/2$.

We conclude that inside the expectation value we may make the replacements
\eqn\rep{\eqalign{
\psi_j(t) \pb_l(u) & \to G(t,u;\f, b)_{jl}  \cr
\psi_j(0) \pb_l(0) & \to (1/2) \d_{jl}.
}}

\appendix B  {Alternative proof of Ward identities and conserved currents}

	We write the action \eact\ in the form $I = \int dt d^dx \ {\cal
L}_{d+1}$.  We
suppose that the Lagrangian density ${\cal L}_{d+1}$  is invariant under the
infinitesimal transformation
$\d \phi =\e \l \phi$, $\d\psi=\e \l \psi $, $\d \pb = - \e \pb \l$, and
$\d b = - \e b \l$, where $\e$ is an infinitesimal constant and $\l$ is a
numerical matrix with $\Tr \l = 0$.   If instead $\e = \e(x,t)$ is an
infinitesimal function in $d+1$ dimensions, then
according to Noether's theorem, the Lagrangian density changes according to
\eqn\emma{\eqalign{
\d {\cal L}_{d+1} = { { \p {\cal L}_{d+1} } \over {\p \ \p_M A_i} } (\l
A)_i \p_M \e
	\equiv K_M \p_M \e \ ,
}}
under this transformation, where $A_i = (\f, \psi,
\pb, b)$
represents the set of all fields, and
$\d A_i = (\l A)_i \e$ their
transform.  Here $K_M$, for $M = 1,... (d+1)$, is the Noether current of the
$d+1$-dimensional topological action \eact.  For this action,
the time-component of the Noether current has the particularly simple
expression
\eqn\kfive{\eqalign{
K_{d+1} = s(\pb \l \f) = b \l \f - \pb \l \psi \ .
}}
Let $\O[\phi]$ be an observable.  Upon making the  preceding infinitesimal
change of
variables  for arbitrary $\e(x, t)$  in the functional integral,
$\langle \O \rangle|_{d+1} = \int {\cal D}A \  \O \exp I$
we obtain the
$(d+1)$-dimensional Ward identity corresponding to the above symmetry,
\eqn\emmaf{\eqalign{
0 & = \langle \Big [  {{\delta^{d+1}  \O  }\over {\delta  \phi(x,t)}} \l
\f(x, t)
- \O \partial _M K_M \Big]  \rangle_{d+1}   \cr
& = \int {\cal D}A_i(x,t) \
\Big [  {{\delta^{d+1}  \O  }\over {\delta  \phi(x,t)}} \l \f(x, t)
- \O \partial _M K_M \Big]
\exp I \ ,
}}
where ${\cal D}A_i(x,t) \equiv {\cal D}\phi_(x,t)
{\cal D}\Psi_(x,t)
{\cal D}\bar\Psi_(x,t) {\cal D}b_(x,t)$,
and ${{\d^{d+1}  \O  }\over{\d \phi(x,t)}}$
designates the functional derivative in
$(d+1)$ dimensions.

	We specialize to the case where $\O$ is a physical
observable, namely it is a functional only of
$\f(x,t)$ on the time slice $t = 0$, so
$\O = \O[\f(x,0)] = \O[\f(x)]|_{\f(x) = \f(x,0)}$.   For such functionals one
easily verifies that the functional derivative may be written
\eqn\fnld{\eqalign{
{{\d^{d+1}  \O  }\over {\d  \phi (x,t)}}
= {{\d^d  \O  }\over {\d  \phi(x)}}|_{\f(x) = \f(x,0)} \ \d(t) .
}}
Integrate over $\int_{-\e}^{\e} dt$, and let $\e \to 0$ to obtain
\eqn\emmad{\eqalign{
0=\int
{\cal D}A_i(x,t) \
\Big [  \Big({{\d^d  \O  }\over {\d  \phi(x)}} \l \phi(x)\Big)|_{\f(x) =
\f(x,0)}
- \O \  [K_{d+1}(x,\e) - K_{d+1}(x,-\e)] \Big]
\exp I \ .
}}
The first term is the desired variation of the observable in $d$ dimensions.

	We next manipulate $K_{d+1}(x,t)|_{-\e}^{+\e}$ to obtain the
Noether current in $d$
dimensions.  The second term in \kfive\ contributes
$(\pb \l \psi)(x,\e) - (\pb \l \psi)(x,-\e)$.  As shown in Appendix~A, we
may make the replacement
$\psi_i(t) \pb_j(t) \to \demi \d_{ij}$ inside the expectation
value, so the two ghost terms cancel.
The first term of \kfive\ contributes, $(b \l \f)(x,\e) - (b \l
\f)(x,-\e)$.  We apply the
time-reversal transformation of Sec. 3 to $(b \l \f)(x,-\e)$, under
which
$\f(x, -\e) \to \f(x, \e)$ and
$b(x, -\e) \to - b(x, \e) - { {\d^d S} \over { \d \f(x)} }|_{\f(x) =
\f(x,\e)}$, and which
leaves
$\O[\f(x,0)]$ invariant. This gives
\eqn\emmae{\eqalign{
0=\int
{\cal D}A_i(x,t) \
\Big [ & \ \Big({{\d^d  \O  }\over {\d  \phi(x)}} \l \phi(x)\Big)|_{\f(x) =
\f(x,0)}  \cr
 & - \O \ 2 (b \l \f)(x,\e)
	- \O \Big( { {\d^d S} \over { \d \f(x)} } \l \f(x)\Big)|_{\f(x) =
\f(x,\e)} \Big]
\exp I \ .
}}
	The term involving $b = s\pb$ gives a vanishing contribution.  For
by $s$-invariance of
the action, this term may be replaced by
\eqn\sexpa{\eqalign{2 \pb(x,\e) s[\l \f(x,\e) \O]
=  2 (\pb \l \psi)(x,\e) \O + 2  (\pb \l \f)(x,\e)
\int d^d y{{\d^d  \O  }\over {\d  \phi(y)}}|_{\f(y) = \f(y,0)}\psi(y,0) \ .
}}
As shown in Appendix A, we may make the
replacement $\psi(y,0) \pb(x,\e) \to G(y,0; x,\e; \f, b)$
inside the expectation-value, where $G(y,0; x,\e; \f, b)$ is the
ghost propagator in fixed external $\f$ and $b$ fields.  However, as is
also shown in
Appendix A, this propagator is retarded, and therefore vanishes for
positive $\e$.  As noted there, we may replace $(\pb \l \psi)(x,\e)$ by
${\rm const} \Tr \l = 0$ .

	Finally, we suppose that the transformation $\d \f = \e \l \f$ for
constant $\e$ is a
symmetry transformation of the $d$-dimensional action $S$.  Then by the
$d$-dimensional version of \emma, we have
$\big[{ {\d^d S} \over { \d \f(x)} }\l \f(x)\big]|_{\f(x) = \f(x,\e)}  = -
\p_\m
j_\m(x,\e)$,
for $\m = 1,... d$, where $j_\m$ is the $d$-dimensional Noether current of the
Euclidean action $S$.  Thus the Ward identity reads
\eqn\emmag{\eqalign{
0=\int
{\cal D}A_i(x,t) \
\Big [  \Big({{\d^d  \O  }\over {\d  \phi(x)}} \l \phi(x)\Big)|_{\f(x) =
\f(x,0)}
+ \p_\m j_\m(x,\e) \ \O[\f(x)]|_{\f(x) = \f(x,0)}  \Big]
\exp I \ .
}}
On taking the limit $\e \to 0$, one recovers the $d$-dimensional Ward identity,
\eqn\dward{\eqalign{
0 = \langle
\Big (  {{\d^d  \O  }\over {\d  \phi(x)}} \l \phi(x)
+ \O[\f(x)]  \p_\m j_\m(x)  \Big)|_{\f(x) = \f(x,0)}
\rangle_{d+1} \ ,
}}
in the $(d+1)$-dimensional topological theory.  [The signs of
\emmaf\ and \dward\ are consistent because the weights are $\exp(+I)$ and
$\exp(-S)$.]


\footatend\vfill\supereject\immediate\closeout\rfile\writestoppt
\baselineskip=14pt\centerline{{\bf References}}\bigskip{\frenchspacing%
\parindent=20pt\escapechar=` \input refs.tmp\vfill\eject}\nonfrenchspacing

\newsec{Figure captions}

Figure 1:  Every vertex has one emerging $b$-line.  One may contunously
follow $(D_0)_{\f b}$ propagators from every vertex of a truncated diagram
to an external $b$-line.  The arrows follow the direction of increasing time.

Figure 2:  A cut diagram, with one $(D_0)_{\f b}$ intermediate line.

\vfill
\eject
\epsfxsize=4in
\epsfbox{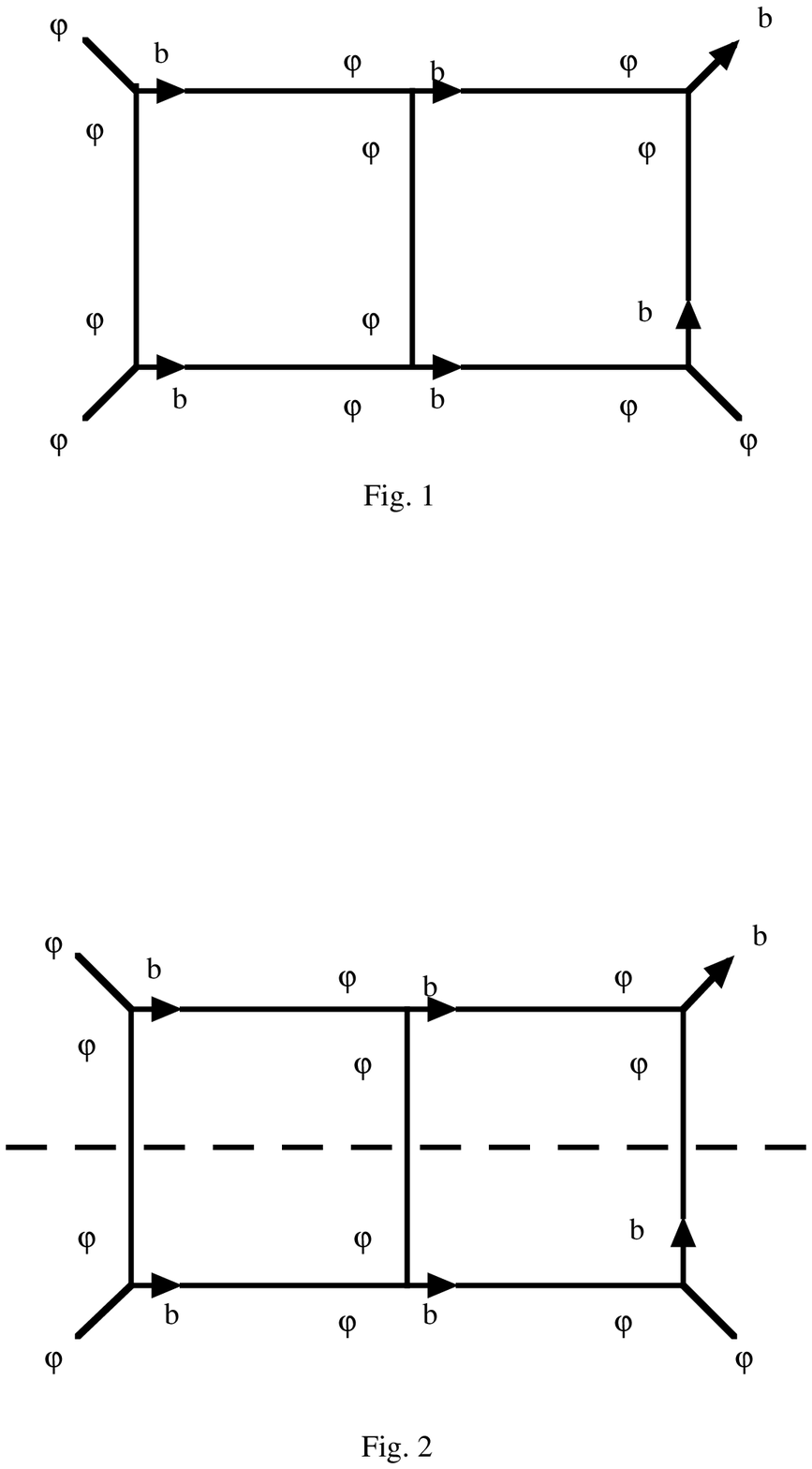}
%

\bye